\documentclass[twocolumn, aps, prd, superscriptaddress,floatfix]{revtex4-1}

\usepackage{multirow}
\usepackage{amsmath,amssymb}
\usepackage{graphicx,bm}
\usepackage{slashed}
\usepackage{epstopdf}
\usepackage{ulem} 
\usepackage[usenames]{color}
\usepackage{float}
\usepackage{braket}
\usepackage{appendix}
\usepackage{hyperref}

\renewcommand\sout{\bgroup \color{red} \ULdepth=-.5ex \ULset}

\begin{document}

\title{Octaquarks in a simple chromomagnetic model}

\author{Aaron Park}
\email{aaron.park@yonsei.ac.kr}\affiliation{Department of Physics and Institute of Physics and Applied Physics, Yonsei University, Seoul 03722, Korea}

\date{\today}
\begin{abstract}
In this work, we study the octaquark, which is composed of two baryons and one meson using a simple chromomagnetic model. First, we construct the wave function of the octaquark satisfying the Pauli exclusion principle in the flavor SU(3) breaking case. In order to calculate the binding energy, we consider the lowest threshold by taking into account the decay products. By calculating the color-spin interaction of octaquarks for all possible quantum numbers, we determine several candidates for compact octaquarks.
\end{abstract}

\maketitle

\section{Introduction}

The study of exotic hadrons \cite{Jaffe:1976ig,Swanson:2006st,Nielsen:2009uh,Brambilla:2019esw,Liu:2019zoy,Richard:2016eis,Ali:2017jda,Hosaka:2016pey,ExHIC:2017smd}, which are particles that do not fit into the conventional categories of baryons and mesons, has attracted significant attention in hadron physics. So far, research has mainly focused on tetraquarks, pentaquarks, and dibaryons, which consist of two hadrons. However, according to the quark model, exotic hadrons consisting of three or more hadrons are also possible. 

First, there is a heptaquark, which consists of one baryon and two mesons: $q^5 \overline{q}^2$ \cite{Bicudo:2003rw,Park:2017mdp,Luo:2022cun}. Next, we can consider the octaquark, which can come in two types: $q^7 \overline{q}$(or $\overline{q}^7 q$) and $q^4 \overline{q}^4$. In the case of the former, it is composed of two baryons and one meson, while in the latter case, it can be composed of four mesons, or one baryon, one antibaryon, and one meson. Next, in the case of a tribaryon, which is made up of three baryons, some research has already been conducted, as it is related to the three-body nuclear force\cite{Maezawa:2004va,Garcilazo:2016ylj,Garcilazo:2016gkj,Park:2018ukx,Park:2019jff,Park:2020qpd}. However, for the heptaquark and octaquark, relatively a few research has been done \cite{Bicudo:2003rw,Park:2017mdp,Luo:2022cun,Gordillo:2023tnz}. Research on exotic nuclei, such as $\overline{D}NN$, has occasionally been conducted \cite{Yasui:2009bz}, however, extensive studies on the existence of compact octaquarks have not yet been carried out. Particularly, heptaquarks and octaquarks contain antiquarks, making the calculations more complex from the perspective of multiquark configurations, and thus, they are more challenging to study.

In this paper, we investigate whether the octaquark, which is composed of two baryons and one meson, can exist as a compact multiquark state using a simple chromomagnetic interaction. First, we construct the wave function of the octaquark satisfying the Pauli exclusion principle in the flavor SU(3) breaking case. Then, since the octaquark can decay into two baryons and one meson, we calculate the binding energy of octaquarks from the perspective of the color-spin interaction, taking into account the lowest threshold.

This paper is organized as follows. In Sec.\ref{Color-spin interaction}, we introduce the color-spin interaction and present the formula used to calculate the binding energy in this study. In Sec.\ref{Color, Spin and flavor states of the octaquark}, we construct the wave function of the octaquark satisfying the Pauli exclusion princple in the flavor SU(3) breaking case. In Sec.\ref{Results}, we present the binding energies of octaquarks considering the lowest decay channels. Sec.\ref{Summary} devoted to summary and concluding remarks.

\section{Color-spin interaction}
\label{Color-spin interaction}

In this study, we use the color-spin interaction to calculate the stability of octaquarks. The color-spin interaction plays a crucial role at short distances and is therefore a key potential for computing compact multiquark states and studying hadron-hadron interactions at short range \cite{DeRujula:1975qlm,Oka:1981ri,Park:2019bsz}.

Addtionally, we assume that all quarks occupy the same spatial configuration. This implies that the spatial part of the potential is the same for any pair, and the relative strength is determined by the color and/or spin structures. The factor for the color-spin interaction is defined as follows.
\begin{align}
V_{CS}&=-A\sum_{i<j}^n \frac{1}{m_i m_j}\lambda^c_i \lambda^c_j \sigma_i \cdot \sigma_j \nonumber \\
&\equiv \frac{A}{m_u^2} H_{CS},
\label{color-spin}
\end{align}
where $\lambda^c_i$, $m_i$, $m_u$ are respectively the color SU(3) Gell-Mann matrices, the constituent quark mass of the $i$'th quark, and the constutient quark mass of $u,d$ quarks. The overall constant $A$ is taken to be a constant determined from the mass difference between a proton and a delta isobar. Since $-\sum \lambda_i^c \lambda_j^c \sigma_i \cdot \sigma_j$ for a proton and a delta are -8 and 8, respectively, so we extract the value $A/m_u^2=18.125$ MeV. For flavor SU(3) symmetric case, color-spin factor $H_{CS}$ can be easily obtained by the following formula.
\begin{align}
  H_{CS}&=-\sum_{i<j}^n \lambda^c_i \lambda^c_j  \sigma_i \cdot \sigma_j \nonumber \\
  &= n(n-10) + \frac{4}{3}S(S+1)+4C_F+2C_C, \nonumber\\
  4C_F &= \frac{4}{3}(p_1^2+p_2^2+3p_1+3p_2+p_1 p_2),
\label{CSF-1}
\end{align}
where $C_F$ and $C_C$ are the first kind of the Casimir operators of flavor and color SU(3), respectively, and $p_i$ is the number of columns containing $i$ boxes in a column in Young diagram.

\section{Color, Spin and flavor states of the octaquark}
\label{Color, Spin and flavor states of the octaquark}

\subsection{Color basis function}

Octaquarks are composed of seven quarks and one antiquark. Since the antiquark is color antitriplet, the remaining seven quarks should be color triplet. Therefore, we can represent the color state of octaquark using the Young diagram as follows.

\begin{align}
    \left(
    \begin{tabular}{|c|c|c|}
         \hline
         \quad \quad & \quad \quad & \quad \quad  \\
         \hline
         \quad \quad & \quad \quad \\
         \cline{1-2}
         \quad \quad & \quad \quad \\
         \cline{1-2}
    \end{tabular},
    \begin{tabular}{|c|}
         \hline
         $\hspace{0.07cm} \overline{8}\hspace{0.07cm}$  \\
         \hline
    \end{tabular}\right)_C
\end{align}

Here, I represent the color state of 7-quark and an antiquark separately because there is no permutation symmetry between them. Since the antitriplet part of antiquark is common, I represent the 21 color basis only using seven-quark part below. For convenience, we omit the ket notation in the Young-Yamanouchi symbol. \\

\noindent $C_1 = 
\begin{tabular}{|c|c|c|}
  \hline
  1 & 2 & 3 \\
  \hline
  4 & 5 \\
  \cline{1-2}
  6 & 7 \\
  \cline{1-2}
\end{tabular}, 
C_2 = 
\begin{tabular}{|c|c|c|}
  \hline
  1 & 2 & 4 \\
  \hline
  3 & 5 \\
  \cline{1-2}
  6 & 7 \\
  \cline{1-2}
\end{tabular},
C_3 = 
\begin{tabular}{|c|c|c|}
  \hline
  1 & 3 & 4 \\
  \hline
  2 & 5 \\
  \cline{1-2}
  6 & 7 \\
  \cline{1-2}
\end{tabular},\\
C_4 = 
\begin{tabular}{|c|c|c|}
  \hline
  1 & 2 & 5 \\
  \hline
  3 & 4 \\
  \cline{1-2}
  6 & 7 \\
  \cline{1-2}
\end{tabular},
C_5 = 
\begin{tabular}{|c|c|c|}
  \hline
  1 & 3 & 5 \\
  \hline
  2 & 4 \\
  \cline{1-2}
  6 & 7 \\
  \cline{1-2}
\end{tabular},
C_6 = 
\begin{tabular}{|c|c|c|}
  \hline
  1 & 2 & 3 \\
  \hline
  4 & 6 \\
  \cline{1-2}
  5 & 7 \\
  \cline{1-2}
\end{tabular},\\
C_7 = 
\begin{tabular}{|c|c|c|}
  \hline
  1 & 2 & 4 \\
  \hline
  3 & 6 \\
  \cline{1-2}
  5 & 7 \\
  \cline{1-2}
\end{tabular},
C_8 = 
\begin{tabular}{|c|c|c|}
  \hline
  1 & 3 & 4 \\
  \hline
  2 & 6 \\
  \cline{1-2}
  5 & 7 \\
  \cline{1-2}
\end{tabular},
C_9 = 
\begin{tabular}{|c|c|c|}
  \hline
  1 & 2 & 5 \\
  \hline
  3 & 6 \\
  \cline{1-2}
  4 & 7 \\
  \cline{1-2}
\end{tabular},\\
C_{10} = 
\begin{tabular}{|c|c|c|}
  \hline
  1 & 3 & 5 \\
  \hline
  2 & 6 \\
  \cline{1-2}
  4 & 7 \\
  \cline{1-2}
\end{tabular},
C_{11} = 
\begin{tabular}{|c|c|c|}
  \hline
  1 & 4 & 5 \\
  \hline
  2 & 6 \\
  \cline{1-2}
  3 & 7 \\
  \cline{1-2}
\end{tabular},
C_{12} = 
\begin{tabular}{|c|c|c|}
  \hline
  1 & 2 & 6 \\
  \hline
  3 & 4 \\
  \cline{1-2}
  5 & 7 \\
  \cline{1-2}
\end{tabular},\\
C_{13} = 
\begin{tabular}{|c|c|c|}
  \hline
  1 & 3 & 6 \\
  \hline
  2 & 4 \\
  \cline{1-2}
  5 & 7 \\
  \cline{1-2}
\end{tabular},
C_{14} = 
\begin{tabular}{|c|c|c|}
  \hline
  1 & 2 & 6 \\
  \hline
  3 & 5 \\
  \cline{1-2}
  4 & 7 \\
  \cline{1-2}
\end{tabular},
C_{15} = 
\begin{tabular}{|c|c|c|}
  \hline
  1 & 3 & 6 \\
  \hline
  2 & 5 \\
  \cline{1-2}
  4 & 7 \\
  \cline{1-2}
\end{tabular},\\
C_{16} = 
\begin{tabular}{|c|c|c|}
  \hline
  1 & 4 & 6 \\
  \hline
  2 & 5 \\
  \cline{1-2}
  3 & 7 \\
  \cline{1-2}
\end{tabular},
C_{17} = 
\begin{tabular}{|c|c|c|}
  \hline
  1 & 2 & 7 \\
  \hline
  3 & 4 \\
  \cline{1-2}
  5 & 6 \\
  \cline{1-2}
\end{tabular},
C_{18} = 
\begin{tabular}{|c|c|c|}
  \hline
  1 & 3 & 7 \\
  \hline
  2 & 4 \\
  \cline{1-2}
  5 & 6 \\
  \cline{1-2}
\end{tabular},\\
C_{19} = 
\begin{tabular}{|c|c|c|}
  \hline
  1 & 2 & 7 \\
  \hline
  3 & 5 \\
  \cline{1-2}
  4 & 6 \\
  \cline{1-2}
\end{tabular},
C_{20} = 
\begin{tabular}{|c|c|c|}
  \hline
  1 & 3 & 7 \\
  \hline
  2 & 5 \\
  \cline{1-2}
  4 & 6 \\
  \cline{1-2}
\end{tabular},
C_{21} = 
\begin{tabular}{|c|c|c|}
  \hline
  1 & 4 & 7 \\
  \hline
  2 & 5 \\
  \cline{1-2}
  3 & 6 \\
  \cline{1-2}
\end{tabular}$.\\

We can also represent the color basis in a tensor form. From $C_{21}$, the tensor form for the remaining bases can be obtained using permutations. A simple example of how to calculate $C_{20}$ is as follows.
\begin{align}
    & C_{21} \nonumber \\
    &=\frac{1}{6\sqrt{3}} \varepsilon_{ijk}q^i(1)q^j(2)q^k(3) \varepsilon_{lmn}q^l(4)q^m(5)q^n(6) q^p(7)\overline{q}^p(8) \nonumber\\
    &\equiv \Psi_C(1,2,3,4,5,6,7,8), \nonumber\\
    & C_{20} \nonumber \\
    &= \frac{3}{2\sqrt{2}} \left( \Psi_C(1,2,4,3,5,6,7,8)-\frac{1}{3}\Psi_C(1,2,3,4,5,6,7,8) \right). \nonumber
\end{align}

\subsection{Spin basis function}

Here, we represent the spin state of octaquark in a Young diagram form. Details of all possible Young-Yamanouchi bases are provided in the appendix \ref{Spin basis of octaquark}.

\begin{itemize}
\item $S=4$:
$\begin{tabular}{|c|c|c|c|c|c|c|c|}
    \hline
    \quad \quad & \quad \quad & \quad \quad & \quad \quad & \quad \quad & \quad \quad & \quad \quad &  \quad \quad  \\
    \hline
\end{tabular}$
\item $S=3$:
$\begin{tabular}{|c|c|c|c|c|c|c|}
    \hline
    \quad \quad & \quad \quad & \quad \quad & \quad \quad & \quad \quad & \quad \quad & \quad \quad \\
    \hline
    \quad \quad \\
    \cline{1-1}
\end{tabular}$
\item $S=2$:
$\begin{tabular}{|c|c|c|c|c|c|}
    \hline
    \quad \quad & \quad \quad & \quad \quad & \quad \quad & \quad \quad & \quad \quad  \\
    \hline
    \quad \quad & \quad \quad \\
    \cline{1-2}
\end{tabular}$
\item $S=1$:
$\begin{tabular}{|c|c|c|c|c|}
    \hline
    \quad \quad & \quad \quad & \quad \quad & \quad \quad & \quad \quad   \\
    \hline
    \quad \quad & \quad \quad & \quad \quad \\
    \cline{1-3}
\end{tabular}$
\item $S=0$:
$\begin{tabular}{|c|c|c|c|}
    \hline
    \quad \quad & \quad \quad & \quad \quad & \quad \quad  \\
    \hline
    \quad \quad & \quad \quad & \quad \quad & \quad \quad \\
    \cline{1-4}
\end{tabular}$
\end{itemize}

\subsection{Flavor basis function}

In this study, we assume that the orbital part of the wave function of octaquarks are totally symmetric. Then, the remaining part of the wave function which is color $\otimes$ spin $\otimes$ flavor state should be antisymmetric to be satisfied with the Pauli principle. There is no permuation symmetry between quarks and antiquarks, but we can utilize the symmetry among the seven quarks. Since the color state of $q^7$ is triplet, which is [322] in terms of Young diagram, the flavor $\otimes$ spin coupling state of $q^7$ should be its conjugate, which is [331] to satisfy the Pauli exclusion principle. Then, we can classify the possible flavor and spin states using flavor SU(3), spin SU(2) groups, and Clebsch-Gordan series of $S_7$ symmetric group as follows.\\

$[331]_{FS}=[61]_F \otimes [43]_S + [52]_F \otimes [52]_S + [52]_F \otimes [43]_S + [511]_F \otimes [52]_S + [511]_F \otimes [43]_S + [43]_F \otimes [61]_S + [43]_F \otimes [52]_S + [43]_F \otimes [43]_S + [421]_F \otimes [61]_S + [421]_F \otimes [52]_{S(m=2)} + [421]_F \otimes [43]_{S(m=2)} + [331]_F \otimes [7]_S + [331]_F \otimes [61]_S + [331]_F \otimes [52]_{S(m=2)} + [331]_F \otimes [43]_S + [322]_F \otimes [61]_S + [322]_F \otimes [52]_S + [322]_F \otimes [43]_S$.\\

Here, $m$ means multiplicity. There are seven possible flavor states for $q^7$ of octaquark as follows.\\

$\begin{tabular}{|c|c|c|c}
  \cline{1-3}
  \quad \quad & \quad \quad & \quad \quad  \\
  \cline{1-3}
  \quad \quad & \quad \quad  \\
  \cline{1-2}
  \quad \quad & \quad \quad  \\
  \cline{1-2}
  \multicolumn{4}{c}{$\mathbf{3}(S=\frac{1}{2},\frac{3}{2},\frac{5}{2})$}
\end{tabular}$,
$\begin{tabular}{|c|c|c|c}
  \cline{1-3}
  \quad \quad & \quad \quad & \quad \quad \\
  \cline{1-3}
  \quad \quad & \quad \quad & \quad \quad \\
  \cline{1-3}
  \quad \quad  \\
  \cline{1-1}
  \multicolumn{4}{l}{$\mathbf{\bar{6}}(S=\frac{1}{2},\frac{3}{2},\frac{5}{2},\frac{7}{2})$}
\end{tabular}$,
$\begin{tabular}{|c|c|c|c|c}
  \cline{1-4}
  \quad \quad & \quad \quad & \quad \quad & \quad \quad \\
  \cline{1-4}
  \quad \quad & \quad \quad  \\
  \cline{1-2}
  \quad \quad  \\
  \cline{1-1}
  \multicolumn{5}{c}{$\mathbf{15}(S=\frac{1}{2},\frac{3}{2},\frac{5}{2})$}
\end{tabular}$,
$\begin{tabular}{|c|c|c|c|c}
  \cline{1-4}
  \quad \quad & \quad \quad & \quad \quad & \quad \quad  \\
  \cline{1-4}
  \quad \quad & \quad \quad & \quad \quad  \\
  \cline{1-3}
  \multicolumn{5}{c}{$\mathbf{24}(S=\frac{1}{2},\frac{3}{2},\frac{5}{2})$}
\end{tabular}$,
$\begin{tabular}{|c|c|c|c|c|}
  \hline
  \quad \quad & \quad \quad & \quad \quad & \quad \quad & \quad \quad  \\
  \hline
  \quad \quad  \\
  \cline{1-1}
  \quad \quad \\
  \cline{1-1}
  \multicolumn{5}{c}{$\mathbf{15'}(S=\frac{1}{2},\frac{3}{2})$}
\end{tabular}$,
$\begin{tabular}{|c|c|c|c|c|}
  \hline
  \quad \quad & \quad \quad & \quad \quad & \quad \quad & \quad \quad \\
  \hline
  \quad \quad & \quad \quad \\
  \cline{1-2}
  \multicolumn{5}{c}{$\mathbf{42}(S=\frac{1}{2},\frac{3}{2})$}
\end{tabular}$,
$\begin{tabular}{|c|c|c|c|c|c|}
  \hline
  \quad \quad & \quad \quad & \quad \quad & \quad \quad & \quad \quad & \quad \quad \\
  \hline
  \quad \quad  \\
  \cline{1-1}
  \multicolumn{6}{c}{$\mathbf{48}(S=\frac{1}{2})$}
\end{tabular}$
\\

In parentheses, we represent the possible spin quantum numbers. Now that we know the flavor and spin states, the color-spin factor of the $q^7$ multiquark can be easily calculated in the flavor SU(3) symmetric limit using the Eq.(\ref{CSF-1}). We represent the color-spin factors of $q^7$ in Table \ref{color-spin factor}.

\begin{table}
\begin{tabular}{|c|c|c|c|c|c|}
  \hline
  \multirow{2}{*}{Flavor} & \multicolumn{4}{|c|}{$-\sum_{i<j} \lambda_i \lambda_j \sigma_i \cdot \sigma_j$} \\
  \cline{2-5}
   & $S=\frac{1}{2}$ & $S=\frac{3}{2}$ & $S=\frac{5}{2}$ & $S=\frac{7}{2}$  \\
  \hline
  $\mathbf{3}$ & $-12$ & $-8$ & $-\frac{4}{3}$ & \\
  \hline
  $\mathbf{\bar{6}}$ &  -4 & 0 & $\frac{20}{3}$ & 16 \\
  \hline
  $\mathbf{15}$ & 4 & 8 & $\frac{44}{3}$ & \\
  \hline
  $\mathbf{24}$ & 16 & 20 & $\frac{80}{3}$ & \\
  \hline
  $\mathbf{15'}$ & 20 & 24 &  & \\
  \hline
  $\mathbf{42}$ & 28 & 32 &  & \\
  \hline
  $\mathbf{48}$ & 48 & & & \\
  \hline
  \end{tabular}
\caption{The color-spin factor of $q^7$ for each flavor in SU(3) flavor symmetric case. The empty cells in the table represent Pauli blocking states.}
\label{color-spin factor}
\end{table}

As we can see in the Table \ref{color-spin factor}, the flavor state which shows most attractive color-spin factor is flavor triplet. Therefore, we can expect that the octaquark, which includes a flavor triplet from the $q^7$ perspective, would be the most attractive state.

The flavor of the octaquark can be expressed using the flavor state of $q^7$ as described above.
\begin{align}
    F^{q^7\overline{Q}}=(F^{q^7},
    \begin{tabular}{|c|}
      \hline
      $\hspace{0.07cm} \overline{8}\hspace{0.07cm}$ \\
      \hline
    \end{tabular}).
\end{align}
The detailed expression of the flavor state of octaquarks for each quantum number is provided in the appendix \ref{Wave function of a octaquark}.


\section{Results}
\label{Results}

Since octaquarks can decay into two baryons and one meson, we calculate the binding energy of octaquarks using the following formula.
\begin{align}
    E_B = V_{CS}^{\text{octaquark}}-V_{CS}^{\text{baryon 1}}- V_{CS}^{\text{baryon 2}}-V_{CS}^{\text{meson}}.
\end{align}
Additionally, since we construct the wave function of octaquarks in the flavor SU(3) broken case, we represent the binding energies of all possible octaquarks as a function of the following parameter.
\begin{align}
    \delta = 1-\frac{m_u}{m_{\overline{Q}}}.
\end{align}
Therefore, when $\delta$ is close to 0, it corresponds to the light quark mass limit, and when $\delta$ is close to 1, it corresponds to the heavy quark mass limit. In this study, we fix the value of $\frac{m_u}{m_s}$ to be $\frac{3}{5}$ for the calculation.

\subsection{$q^7\overline{Q}$}

Here, we fix the position of each quark on $q(1)q(2)q(3)q(4)q(5)q(6)q(7)\overline{Q}(8)$. In this case, the wave function must satisfy the following symmetric property in order to comply with the Pauli exclusion principle: $\{1234567\}8$. The elements inside the curly brackets are antisymmetric. We represent the detailed information about the wave functions in the appendix \ref{Wave function of a octaquark}. 

\begin{figure}[H]
    \centering
    \includegraphics[width=1.0\linewidth]{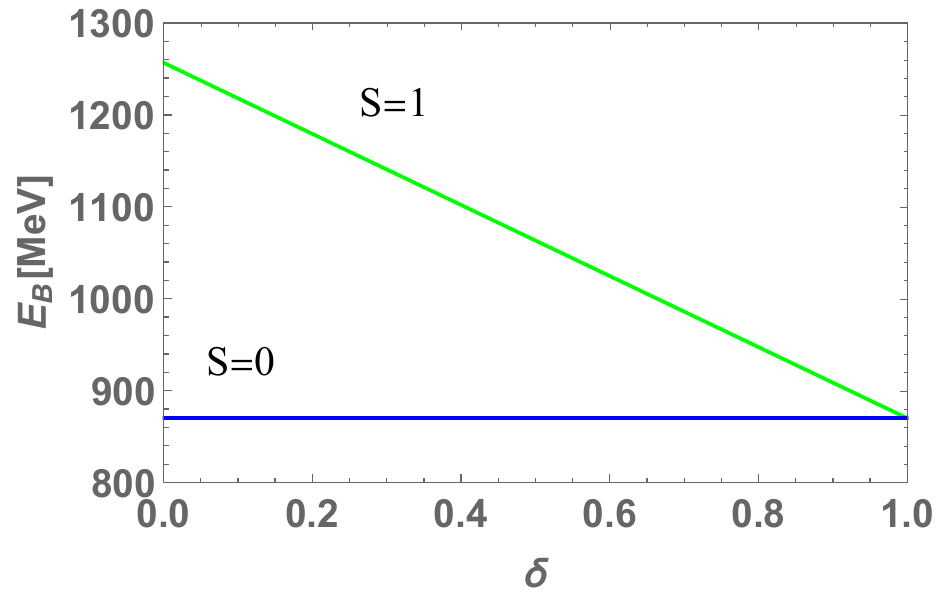}
    \caption{$E_B$ of $q^7\overline{Q}(I=\frac{5}{2})$.}
    \label{qqqqqqqQ(I=2.5)}
\end{figure}

\begin{figure}[H]
    \centering
    \includegraphics[width=1.0\linewidth]{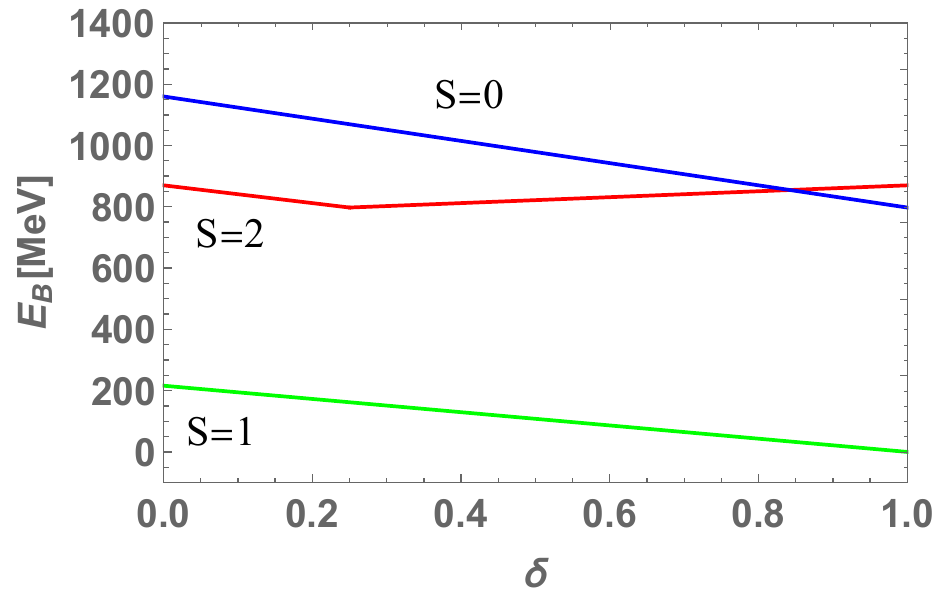}
    \caption{$E_B$ of $q^7\overline{Q}(I=\frac{3}{2})$.}
    \label{qqqqqqqQ(I=1.5)}
\end{figure}

\begin{figure}[H]
    \centering
    \includegraphics[width=1.0\linewidth]{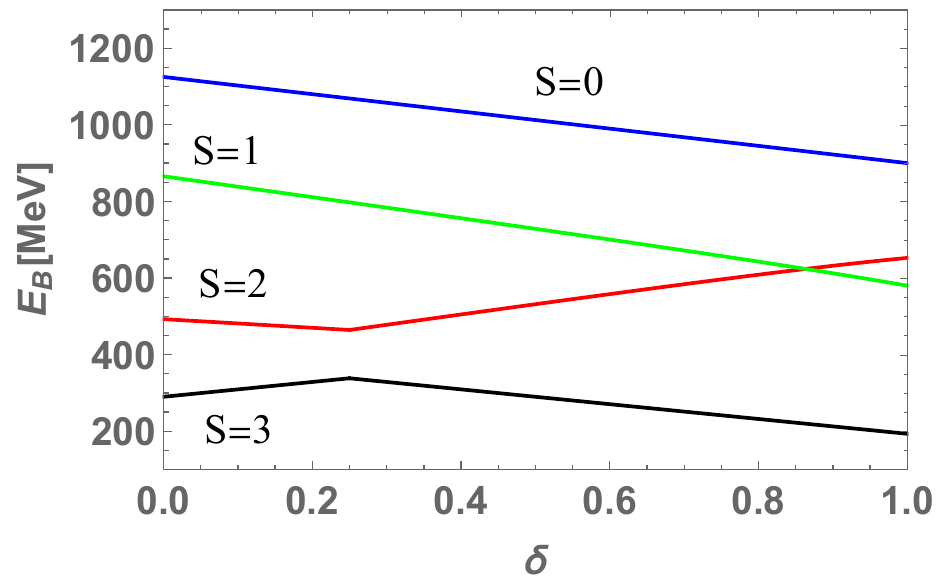}
    \caption{$E_B$ of $q^7\overline{Q}(I=\frac{1}{2})$.}
    \label{qqqqqqqQ(I=0.5)}
\end{figure}

Additionally, since the $I=\frac{7}{2}$ in this case is forbidden by Pauli principle, the possible isospins are $I=\frac{5}{2},\frac{3}{2},\frac{1}{2}$. We plot the binding energies of $q^7 \overline{Q}$ octaquarks as a function of $\delta$ in Fig.\ref{qqqqqqqQ(I=2.5)},\ref{qqqqqqqQ(I=1.5)},\ref{qqqqqqqQ(I=0.5)}. Among these, we can find that $E_B$ of $q^7\overline{Q}(I=\frac{3}{2},S=1)$ state approaches zero in the heavy quark mass limit. The lowest decay channels with respect to this case are $NND$ or $NNB$. Furthermore, since all other cases exhibit very large binding energies, it seems unlikely that a stable compact octaquark can be expected in $q^7 \overline{Q}$ configuration.


\subsection{$q^6s\overline{Q}$}

\begin{figure}[H]
    \centering
    \includegraphics[width=1.0\linewidth]{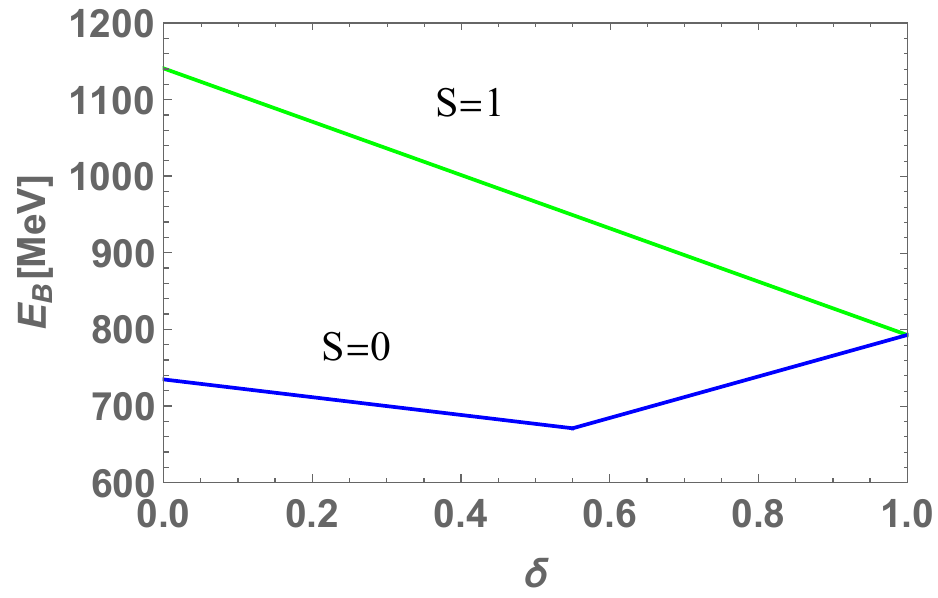}
    \caption{$E_B$ of $q^6s\overline{Q}(I=3)$.}
    \label{qqqqqqsQ(I=3)}
\end{figure}

\begin{figure}[H]
    \centering
    \includegraphics[width=1.0\linewidth]{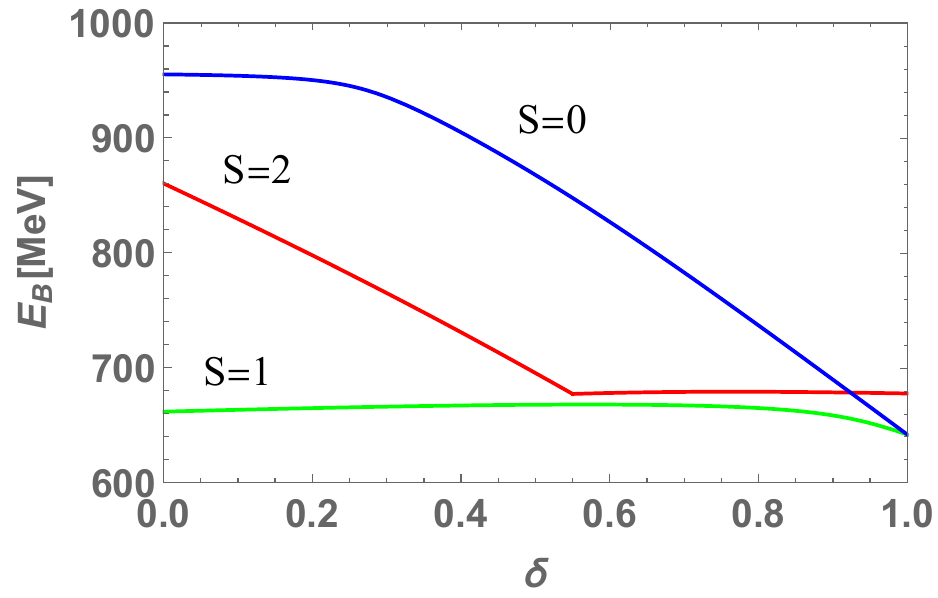}
    \caption{$E_B$ of $q^6s\overline{Q}(I=2)$.}
    \label{qqqqqqsQ(I=2)}
\end{figure}

\begin{figure}[H]
    \centering
    \includegraphics[width=1.0\linewidth]{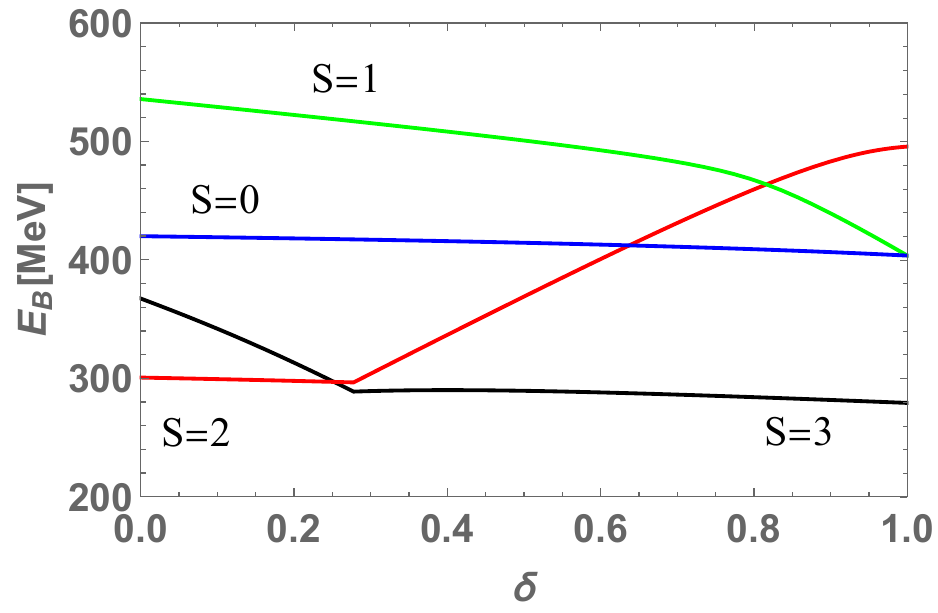}
    \caption{$E_B$ of $q^6s\overline{Q}(I=1)$.}
    \label{qqqqqqsQ(I=1)}
\end{figure}

\begin{figure}[H]
    \centering
    \includegraphics[width=1.0\linewidth]{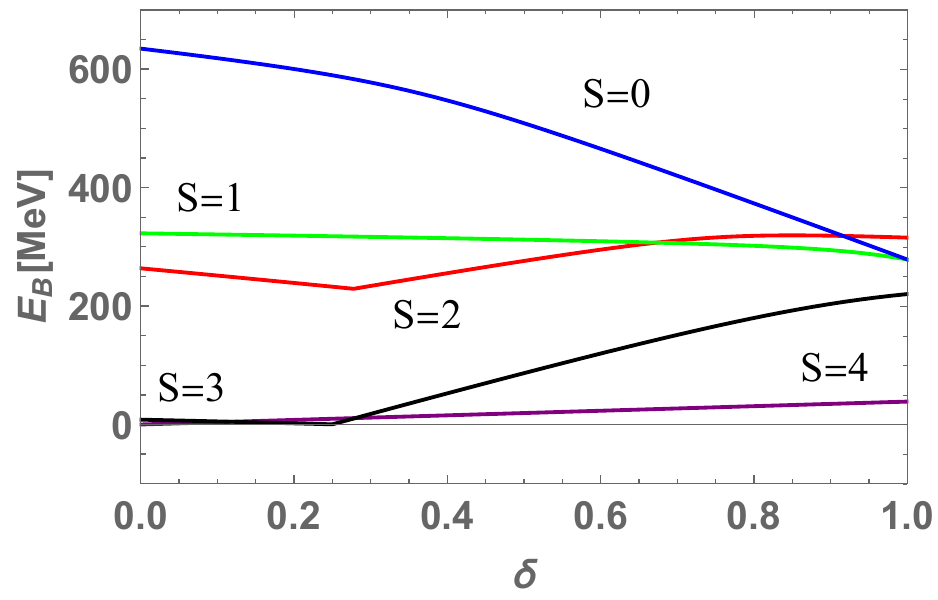}
    \caption{$E_B$ of $q^6s\overline{Q}(I=0)$.}
    \label{qqqqqqsQ(I=0)}
\end{figure}

Here, we fix the position of each quark on $q(1)q(2)q(3)q(4)q(5)q(6)s(7)\overline{Q}(8)$. Then, the wave functions satisfy $\{123456\}78$ symmetry. It should be noted that there is no symmetry between $s$ and $\overline{Q}$. The possible isospins are $I=3,2,1,0$. We plot the binding energies of $q^6s\overline{Q}$ octaquarks in Fig.\ref{qqqqqqsQ(I=3)},\ref{qqqqqqsQ(I=2)},\ref{qqqqqqsQ(I=1)},\ref{qqqqqqsQ(I=0)}. In this case, there are no states with a negative binding energy, but the one closest to 0 are $(I=0,S=4)$ and $(I=0,S=3)$. Here, the binding energy approaches 0 when 
$\delta$ is close to 0, which corresponds to the light quark mass limit. Therefore, in this case, the antiquark becomes $\overline{u}$ or $\overline{d}$, resulting in change of isospin to $\frac{1}{2}$. Then, the lowest thresholds are $\Delta \Delta K^*$ and $\Delta \Delta K$, respectively.


\subsection{$q^5s^2\overline{Q}$}

\begin{figure}[H]
    \centering
    \includegraphics[width=1.0\linewidth]{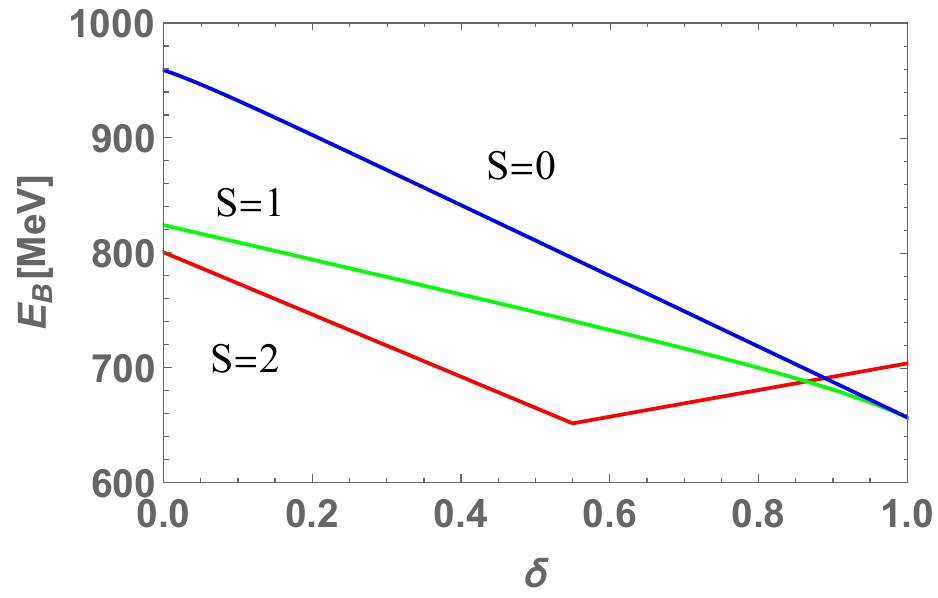}
    \caption{$E_B$ of $q^5s^2\overline{Q}(I=\frac{5}{2})$.}
    \label{qqqqqssQ(I=2.5)}
\end{figure}

\begin{figure}[H]
    \centering
    \includegraphics[width=1.0\linewidth]{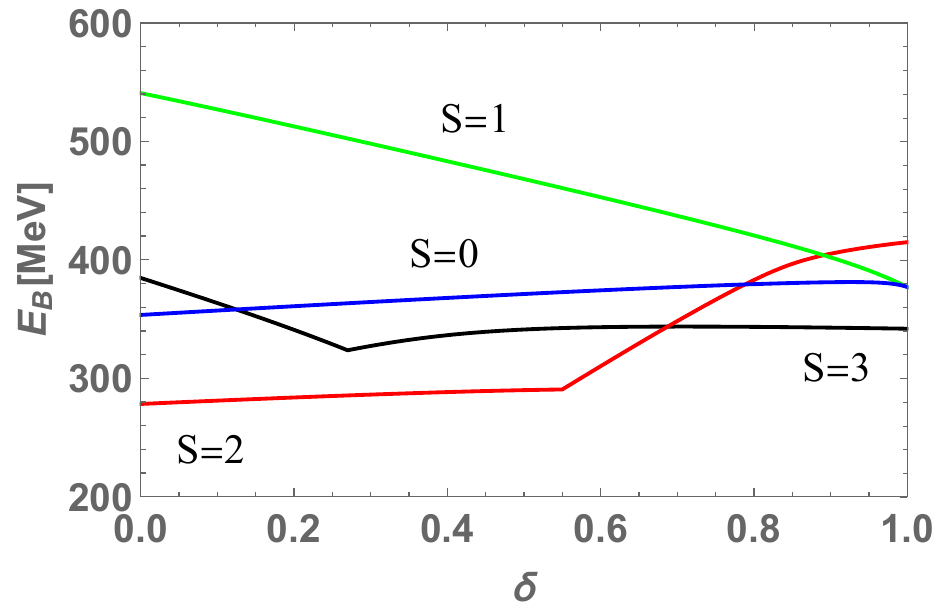}
    \caption{$E_B$ of $q^5s^2\overline{Q}(I=\frac{3}{2})$.}
    \label{qqqqqssQ(I=1.5)}
\end{figure}

\begin{figure}[H]
    \centering
    \includegraphics[width=1.0\linewidth]{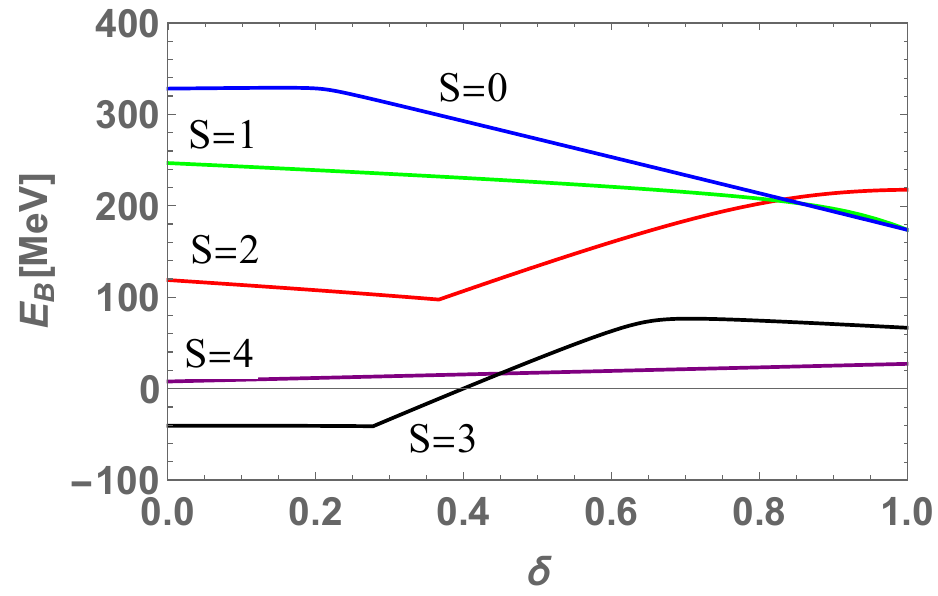}
    \caption{$E_B$ of $q^5s^2\overline{Q}(I=\frac{1}{2})$.}
    \label{qqqqqssQ(I=0.5)}
\end{figure}

Here, we fix the position of each quark on $q(1)q(2)q(3)q(4)q(5)s(6)s(7)\overline{Q}(8)$. Then, the wave functions satisfy $\{12345\}\{67\}8$ symmetry. Note that when there are two or more strange quarks, their symmetry must also be considered. The possible isospins are $I=\frac{5}{2},\frac{3}{2},\frac{1}{2}$. We plot the binding energies of $q^5s^2\overline{Q}$ octaquarks in Fig.\ref{qqqqqssQ(I=2.5)},\ref{qqqqqssQ(I=1.5)},\ref{qqqqqssQ(I=0.5)}. Here, there exists a state with a negative binding energy, which is $(I=\frac{1}{2},S=3)$. In this case, the antiquark becomes $\overline{u}$ or $\overline{d}$, then the isospin becomes 1 or 0. The lowest threshold for this state is $N\Sigma^* K^*$ or $\Delta \Sigma K^*$.


\subsection{$q^4s^3\overline{Q}$}

\begin{figure}[H]
    \centering
    \includegraphics[width=1.0\linewidth]{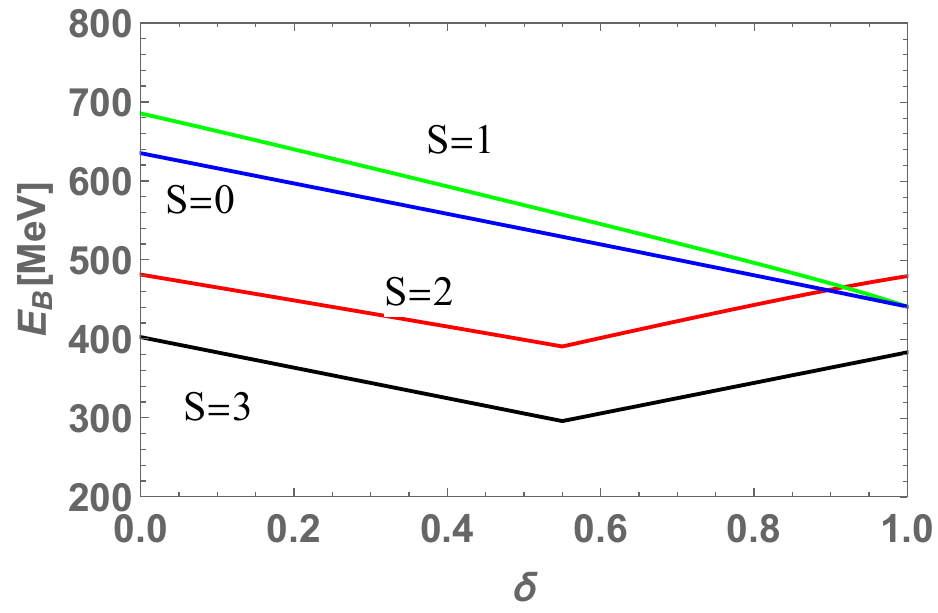}
    \caption{$E_B$ of $q^4s^3\overline{Q}(I=2)$.}
    \label{qqqqsssQ(I=2)}
\end{figure}

\begin{figure}[H]
    \centering
    \includegraphics[width=1.0\linewidth]{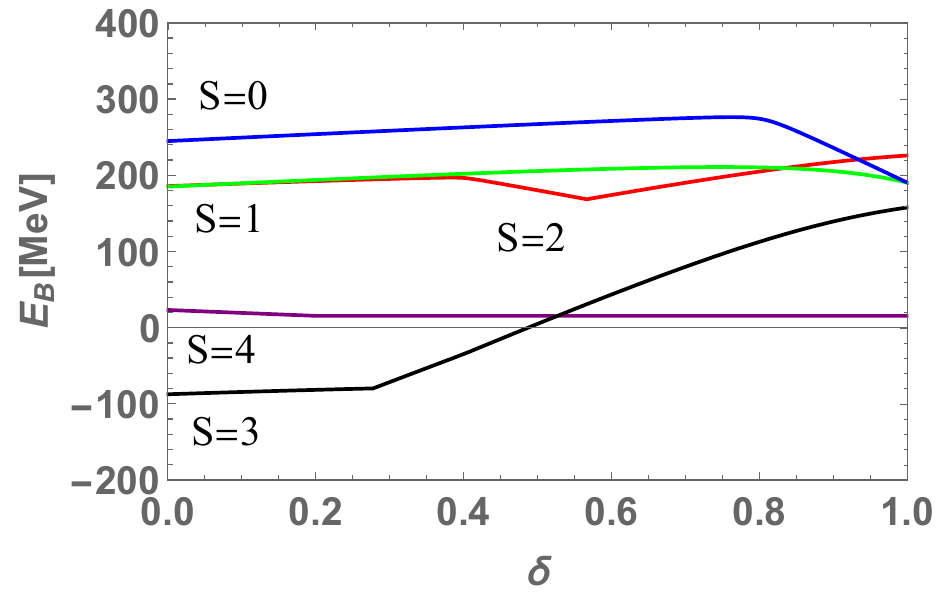}
    \caption{$E_B$ of $q^4s^3\overline{Q}(I=1)$.}
    \label{qqqqsssQ(I=1)}
\end{figure}

\begin{figure}[H]
    \centering
    \includegraphics[width=1.0\linewidth]{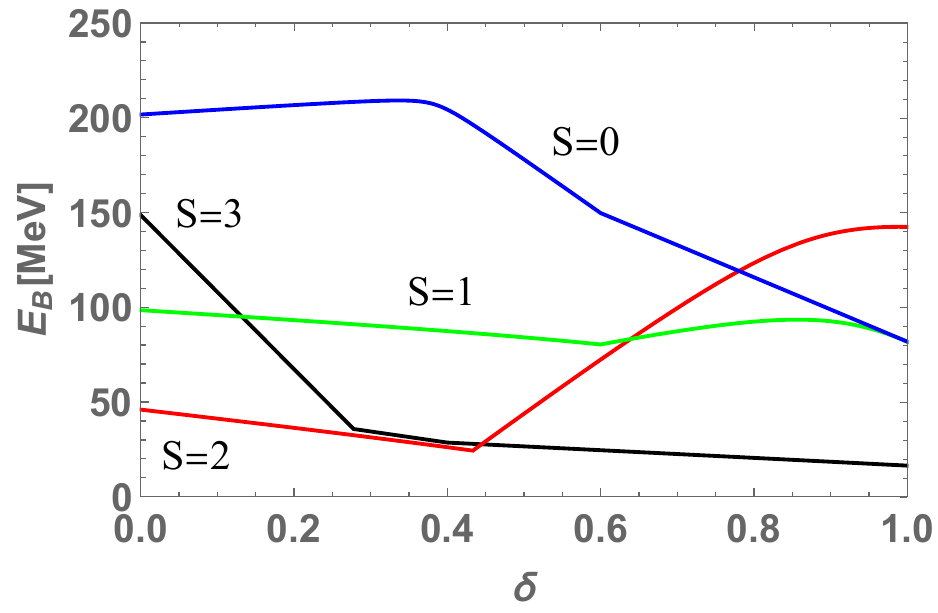}
    \caption{$E_B$ of $q^4s^3\overline{Q}(I=0)$.}
    \label{qqqqsssQ(I=0)}
\end{figure}

Here, we fix the position of each quark on $q(1)q(2)q(3)q(4)s(5)s(6)s(7)\overline{Q}(8)$. Then, the wave functions satisfy $\{1234\}\{567\}8$ symmetry. The possible isospins are $I=2,1,0$. We plot the binding energies of $q^4s^3 \overline{Q}$ octaquarks in Fig.\ref{qqqqsssQ(I=2)},\ref{qqqqsssQ(I=1)},\ref{qqqqsssQ(I=0)}. Among these, there is a state with a negative binding energy, which is $(I=1,S=3)$. In this case, there are two possibilities: $q^4s^3\overline{q}$ or $q^4s^3\overline{s}$. For $q^4s^3\overline{q}$ octaquark, the isospin becomes $\frac{3}{2}$ or $\frac{1}{2}$, and the lowest decay channel is $N\Xi^* K^*$ or $\Delta \Xi K^*$. For $q^4s^3\overline{s}$ octaquark, the lowest decay channel is $N\Xi^* \phi$ or $\Delta \Xi \phi$.  In this study, these states show the greatest potential as a compact octaquark. However, if we consider the additional kinetic energy, it seems unlikely to exist as a ground state, though the possibility of a resonance can be considered. 


\subsection{$q^3s^4\overline{Q}$}

\begin{figure}[H]
    \centering
    \includegraphics[width=1.0\linewidth]{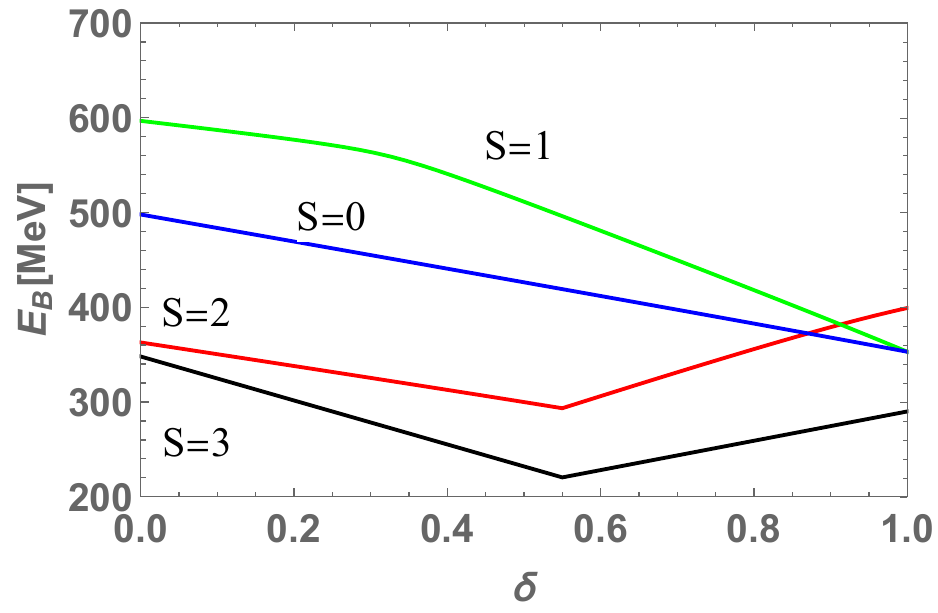}
    \caption{$E_B$ of $q^3s^4\overline{Q}(I=\frac{3}{2})$.}
    \label{qqqssssQ(I=1.5)}
\end{figure}

\begin{figure}[H]
    \centering
    \includegraphics[width=1.0\linewidth]{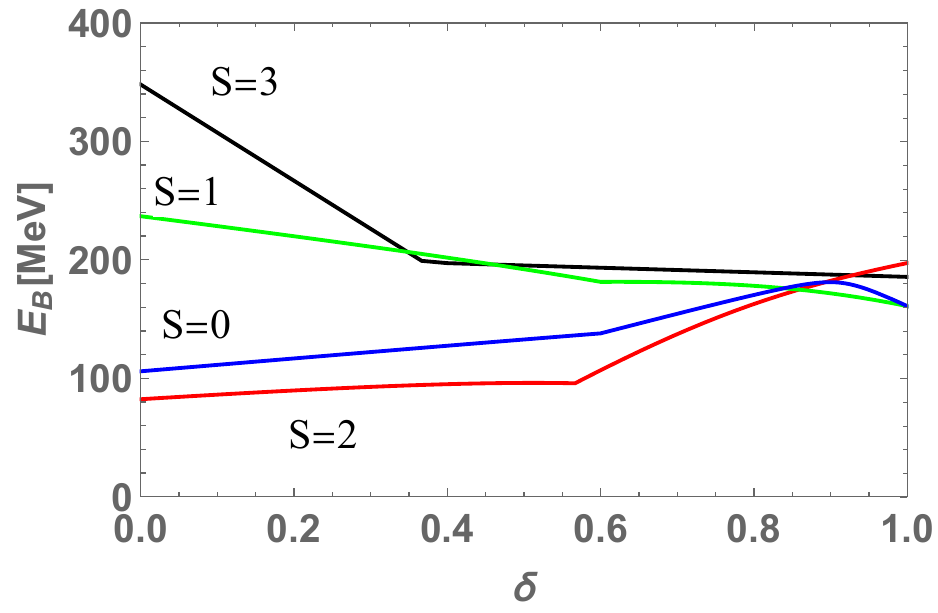}
    \caption{$E_B$ of $q^3s^4\overline{Q}(I=\frac{1}{2})$.}
    \label{qqqssssQ(I=0.5)}
\end{figure}

Here, we fix the position of each quark on $s(1)s(2)s(3)s(4)q(5)q(6)q(7)\overline{Q}(8)$. It should be noted that the strange quarks are positioned at the front for the convenience of calculation. Then, the wave functions satisfy $\{1234\}\{567\}8$ symmetry. The possible isospins are $I=\frac{3}{2},\frac{1}{2}$. We plot the binding energies of $q^3s^4 \overline{Q}$ octaquarks in Fig.\ref{qqqssssQ(I=1.5)},\ref{qqqssssQ(I=0.5)}. There are no states that exhibit a negative binding energy or a binding energy close to 0.


\subsection{$q^2s^5\overline{Q}$}

\begin{figure}[H]
    \centering
    \includegraphics[width=1.0\linewidth]{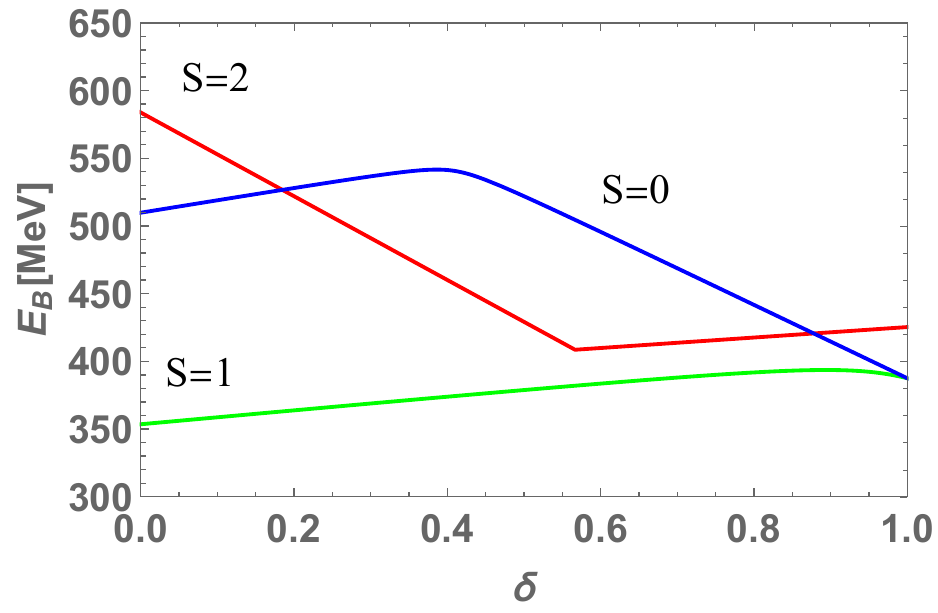}
    \caption{$E_B$ of $q^2s^5\overline{Q}(I=1)$.}
    \label{qqsssssQ(I=1)}
\end{figure}

\begin{figure}[H]
    \centering
    \includegraphics[width=1.0\linewidth]{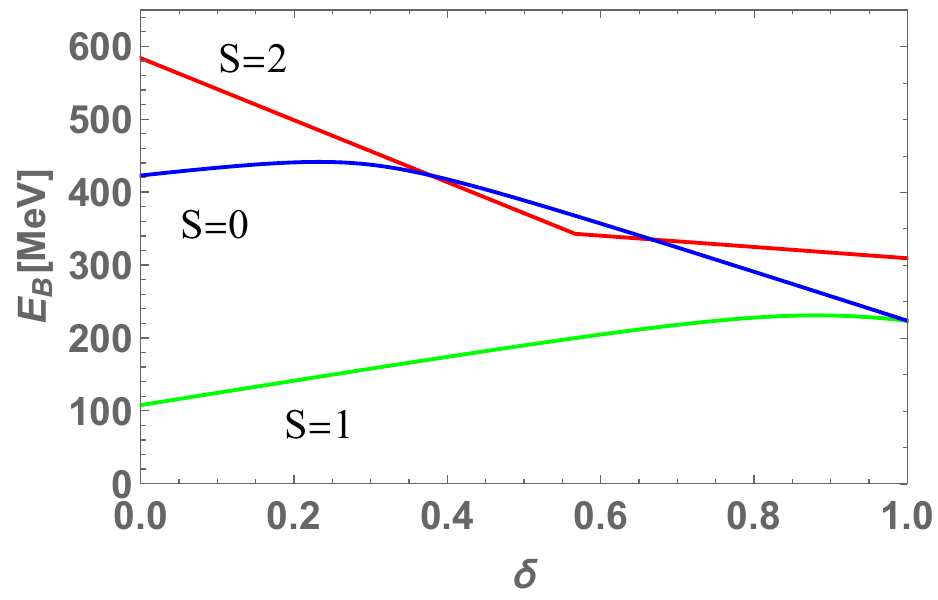}
    \caption{$E_B$ of $q^2s^5\overline{Q}(I=0)$.}
    \label{qqsssssQ(I=0)}
\end{figure}

Here, we fix the position of each quark on $s(1)s(2)s(3)s(4)s(5)q(6)q(7)\overline{Q}(8)$. Then, the wave functions satisfy $\{12345\}\{67\}8$ symmetry. The possible isospins are $I=1,0$. We plot the binding energies of $q^2s^5 \overline{Q}$ octaquarks in Fig.\ref{qqsssssQ(I=1)},\ref{qqsssssQ(I=0)}. There are no states that exhibit a negative binding energy or a binding energy close to 0.


\subsection{$qs^6\overline{Q}$}

\begin{figure}[H]
    \centering
    \includegraphics[width=1.0\linewidth]{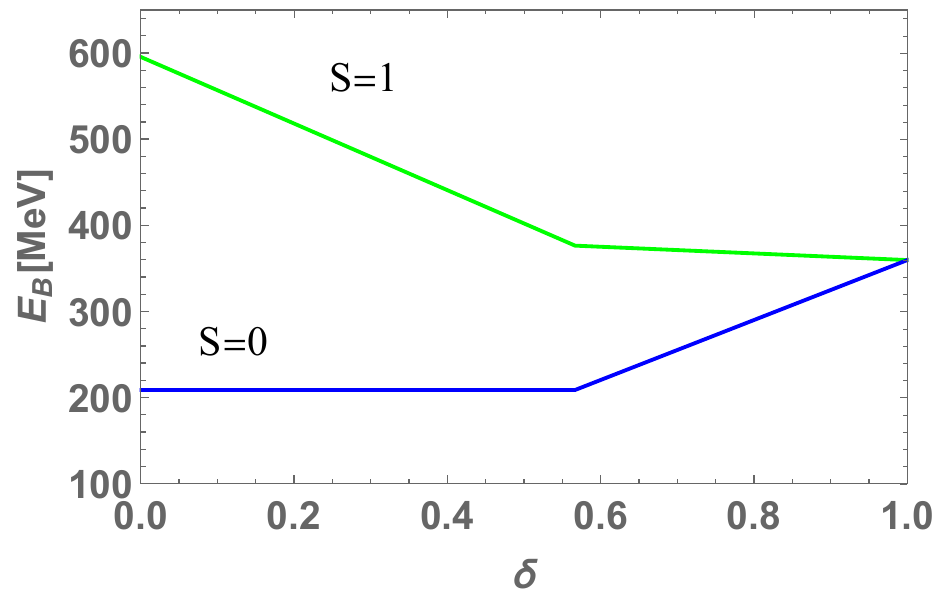}
    \caption{$E_B$ of $qs^6\overline{Q}(I=\frac{1}{2})$.}
    \label{qssssssQ(I=0.5)}
\end{figure}

Here, we fix the position of each quark on $s(1)s(2)s(3)s(4)s(5)s(6)q(7)\overline{Q}(8)$. Then, the wave functions satisfy $\{123456\}78$ symmetry. The possible isospin is $I=\frac{1}{2}$. We plot the binding energies of $qs^6 \overline{Q}$ octaquarks in Fig.\ref{qssssssQ(I=0.5)}. There are no states that exhibit a negative binding energy or a binding energy close to 0.

\section{Summary}
\label{Summary}

In this study, we construct the wave function of octaquarks satisfying the Pauli exclusion principle in the flavor SU(3) breaking case. Using the wave functions of octaquarks and color-spin interaction, we calculate the binding energies of octaquarks considering the lowest decay channels for each quantum number. By calculating the binding energy for all possible quantum numbers and the full range of an antiquark mass, several states with close to zero or negative binding energy were identified.

\begin{itemize}
    \item $q^7\overline{Q}(I=\frac{3}{2},S=1)$: $NND$ or $NNB$
    \item $q^6s\overline{q}(I=\frac{1}{2},S=4)$: $\Delta \Delta K^*$
    \item $q^6s\overline{q}(I=\frac{1}{2},S=3)$: $\Delta \Delta K$
    \item $q^5s^2\overline{q}(I=1,0,S=3)$: $N\Sigma^* K^*$ or $\Delta \Sigma K^*$
    \item $q^4s^3\overline{q}(I=\frac{3}{2},\frac{1}{2},S=3)$: $N\Xi^* K^*$ or $\Delta \Xi K^*$
    \item $q^4s^3\overline{s}(I=1,S=3)$: $N\Xi^* \phi$ or $\Delta \Xi \phi$
\end{itemize}

However, it should be noted that since the binding energies of these states are less than 100 MeV, if the additional kinetic energy is taken into account, the possibility of a compact octaquark's existence becomes significantly lower. 

In this study, we only consider the two-body interaction potential. However, to investigate the compact multiquark configuration, it is necessary to further examine the three-quark potential \cite{Noh:2024nvp}. Additionally, to explore the possibility of a molecular state, the pion exchange potential should also be analyzed \cite{Glozman:1995fu}.

\appendix

\section{Spin basis of octaquark}
\label{Spin basis of octaquark}

In this section, we present the spin basis set of octaquark for $S=4,3,2,1,0$.

\subsection{$S=4$}

$S_1 = 
\begin{tabular}{|c|c|c|c|c|c|c|c|}
    \hline
    1 & 2 & 3 & 4 & 5 & 6 & 7 &  8  \\
    \hline
\end{tabular}$.

\subsection{$S=3$}

\noindent $S_1 = 
\begin{tabular}{|c|c|c|c|c|c|c|}
    \hline
    1 & 2 & 3 & 4 & 5 & 6 & 7 \\
    \hline
    8 \\
    \cline{1-1}
\end{tabular},
S_2 = 
\begin{tabular}{|c|c|c|c|c|c|c|}
    \hline
    1 & 2 & 3 & 4 & 5 & 6 & 8 \\
    \hline
    7 \\
    \cline{1-1}
\end{tabular},\\
S_3 = 
\begin{tabular}{|c|c|c|c|c|c|c|}
    \hline
    1 & 2 & 3 & 4 & 5 & 7 & 8 \\
    \hline
    6 \\
    \cline{1-1}
\end{tabular},
S_4 = 
\begin{tabular}{|c|c|c|c|c|c|c|}
    \hline
    1 & 2 & 3 & 4 & 6 & 7 & 8 \\
    \hline
    5 \\
    \cline{1-1}
\end{tabular},\\
S_5 = 
\begin{tabular}{|c|c|c|c|c|c|c|}
    \hline
    1 & 2 & 3 & 5 & 6 & 7 & 8 \\
    \hline
    4 \\
    \cline{1-1}
\end{tabular},
S_6 = 
\begin{tabular}{|c|c|c|c|c|c|c|}
    \hline
    1 & 2 & 4 & 5 & 6 & 7 & 8 \\
    \hline
    3 \\
    \cline{1-1}
\end{tabular},\\
S_7 = 
\begin{tabular}{|c|c|c|c|c|c|c|}
    \hline
    1 & 3 & 4 & 5 & 6 & 7 & 8 \\
    \hline
    2 \\
    \cline{1-1}
\end{tabular}$.

\subsection{$S=2$}

\noindent$S_1 = 
\begin{tabular}{|c|c|c|c|c|c|}
    \hline
    1 & 2 & 3 & 4 & 5 & 6 \\
    \hline
    7 & 8 \\
    \cline{1-2}
\end{tabular},
S_2 = 
\begin{tabular}{|c|c|c|c|c|c|}
    \hline
    1 & 2 & 3 & 4 & 5 & 7 \\
    \hline
    6 & 8 \\
    \cline{1-2}
\end{tabular},\\
S_3 = 
\begin{tabular}{|c|c|c|c|c|c|}
    \hline
    1 & 2 & 3 & 4 & 6 & 7 \\
    \hline
    5 & 8 \\
    \cline{1-2}
\end{tabular},
S_4 = 
\begin{tabular}{|c|c|c|c|c|c|}
    \hline
    1 & 2 & 3 & 5 & 6 & 7 \\
    \hline
    4 & 8 \\
    \cline{1-2}
\end{tabular},\\
S_5 = 
\begin{tabular}{|c|c|c|c|c|c|}
    \hline
    1 & 2 & 4 & 5 & 6 & 7 \\
    \hline
    3 & 8 \\
    \cline{1-2}
\end{tabular},
S_6 = 
\begin{tabular}{|c|c|c|c|c|c|}
    \hline
    1 & 3 & 4 & 5 & 6 & 7 \\
    \hline
    2 & 8 \\
    \cline{1-2}
\end{tabular},\\
S_7 = 
\begin{tabular}{|c|c|c|c|c|c|}
    \hline
    1 & 2 & 3 & 4 & 5 & 8 \\
    \hline
    6 & 7 \\
    \cline{1-2}
\end{tabular},
S_8 = 
\begin{tabular}{|c|c|c|c|c|c|}
    \hline
    1 & 2 & 3 & 4 & 6 & 8 \\
    \hline
    5 & 7 \\
    \cline{1-2}
\end{tabular},\\
S_9 = 
\begin{tabular}{|c|c|c|c|c|c|}
    \hline
    1 & 2 & 3 & 5 & 6 & 8 \\
    \hline
    4 & 7 \\
    \cline{1-2}
\end{tabular},
S_{10} = 
\begin{tabular}{|c|c|c|c|c|c|}
    \hline
    1 & 2 & 4 & 5 & 6 & 8 \\
    \hline
    3 & 7 \\
    \cline{1-2}
\end{tabular},\\
S_{11} = 
\begin{tabular}{|c|c|c|c|c|c|}
    \hline
    1 & 3 & 4 & 5 & 6 & 8 \\
    \hline
    2 & 7 \\
    \cline{1-2}
\end{tabular},
S_{12} = 
\begin{tabular}{|c|c|c|c|c|c|}
    \hline
    1 & 2 & 3 & 4 & 7 & 8 \\
    \hline
    5 & 6 \\
    \cline{1-2}
\end{tabular},\\
S_{13} = 
\begin{tabular}{|c|c|c|c|c|c|}
    \hline
    1 & 2 & 3 & 5 & 7 & 8 \\
    \hline
    4 & 6 \\
    \cline{1-2}
\end{tabular},
S_{14} = 
\begin{tabular}{|c|c|c|c|c|c|}
    \hline
    1 & 2 & 4 & 5 & 7 & 8 \\
    \hline
    3 & 6 \\
    \cline{1-2}
\end{tabular},\\
S_{15} = 
\begin{tabular}{|c|c|c|c|c|c|}
    \hline
    1 & 3 & 4 & 5 & 7 & 8 \\
    \hline
    2 & 6 \\
    \cline{1-2}
\end{tabular},
S_{16} = 
\begin{tabular}{|c|c|c|c|c|c|}
    \hline
    1 & 2 & 3 & 6 & 7 & 8 \\
    \hline
    4 & 5 \\
    \cline{1-2}
\end{tabular},\\
S_{17} = 
\begin{tabular}{|c|c|c|c|c|c|}
    \hline
    1 & 2 & 4 & 6 & 7 & 8 \\
    \hline
    3 & 5 \\
    \cline{1-2}
\end{tabular},
S_{18} = 
\begin{tabular}{|c|c|c|c|c|c|}
    \hline
    1 & 3 & 4 & 6 & 7 & 8 \\
    \hline
    2 & 5 \\
    \cline{1-2}
\end{tabular},\\
S_{19} = 
\begin{tabular}{|c|c|c|c|c|c|}
    \hline
    1 & 2 & 5 & 6 & 7 & 8 \\
    \hline
    3 & 4 \\
    \cline{1-2}
\end{tabular},
S_{20} = 
\begin{tabular}{|c|c|c|c|c|c|}
    \hline
    1 & 3 & 5 & 6 & 7 & 8 \\
    \hline
    2 & 4 \\
    \cline{1-2}
\end{tabular}$.

\subsection{$S=1$}

\begin{small}
\noindent$S_1 = 
\begin{tabular}{|c|c|c|c|c|}
    \hline
    1 & 2 & 3 & 4 & 5 \\
    \hline
    6 & 7 & 8 \\
    \cline{1-3}
\end{tabular},
S_2 = 
\begin{tabular}{|c|c|c|c|c|}
    \hline
    1 & 2 & 3 & 4 & 6 \\
    \hline
    5 & 7 & 8 \\
    \cline{1-3}
\end{tabular},
S_3 = 
\begin{tabular}{|c|c|c|c|c|}
    \hline
    1 & 2 & 3 & 5 & 6 \\
    \hline
    4 & 7 & 8 \\
    \cline{1-3}
\end{tabular},\\
S_4 = 
\begin{tabular}{|c|c|c|c|c|}
    \hline
    1 & 2 & 4 & 5 & 6 \\
    \hline
    3 & 7 & 8 \\
    \cline{1-3}
\end{tabular},
S_5 = 
\begin{tabular}{|c|c|c|c|c|}
    \hline
    1 & 3 & 4 & 5 & 6 \\
    \hline
    2 & 7 & 8 \\
    \cline{1-3}
\end{tabular},
S_6 = 
\begin{tabular}{|c|c|c|c|c|}
    \hline
    1 & 2 & 3 & 4 & 7 \\
    \hline
    5 & 6 & 8 \\
    \cline{1-3}
\end{tabular},\\
S_7 = 
\begin{tabular}{|c|c|c|c|c|}
    \hline
    1 & 2 & 3 & 5 & 7 \\
    \hline
    4 & 6 & 8 \\
    \cline{1-3}
\end{tabular},
S_8 = 
\begin{tabular}{|c|c|c|c|c|}
    \hline
    1 & 2 & 4 & 5 & 7 \\
    \hline
    3 & 6 & 8 \\
    \cline{1-3}
\end{tabular},
S_9 = 
\begin{tabular}{|c|c|c|c|c|}
    \hline
    1 & 3 & 4 & 5 & 7 \\
    \hline
    2 & 6 & 8 \\
    \cline{1-3}
\end{tabular},\\
S_{10} = 
\begin{tabular}{|c|c|c|c|c|}
    \hline
    1 & 2 & 3 & 6 & 7 \\
    \hline
    4 & 5 & 8 \\
    \cline{1-3}
\end{tabular},
S_{11} = 
\begin{tabular}{|c|c|c|c|c|}
    \hline
    1 & 2 & 4 & 6 & 7 \\
    \hline
    3 & 5 & 8 \\
    \cline{1-3}
\end{tabular},
S_{12} = 
\begin{tabular}{|c|c|c|c|c|}
    \hline
    1 & 3 & 4 & 6 & 7 \\
    \hline
    2 & 5 & 8 \\
    \cline{1-3}
\end{tabular},\\
S_{13} = 
\begin{tabular}{|c|c|c|c|c|}
    \hline
    1 & 2 & 5 & 6 & 7 \\
    \hline
    3 & 4 & 8 \\
    \cline{1-3}
\end{tabular},
S_{14} = 
\begin{tabular}{|c|c|c|c|c|}
    \hline
    1 & 3 & 5 & 6 & 7 \\
    \hline
    2 & 4 & 8 \\
    \cline{1-3}
\end{tabular},
S_{15} = 
\begin{tabular}{|c|c|c|c|c|}
    \hline
    1 & 2 & 3 & 4 & 8 \\
    \hline
    5 & 6 & 7 \\
    \cline{1-3}
\end{tabular},\\
\\
S_{16} = 
\begin{tabular}{|c|c|c|c|c|}
    \hline
    1 & 2 & 3 & 5 & 8 \\
    \hline
    4 & 6 & 7 \\
    \cline{1-3}
\end{tabular},
S_{17} = 
\begin{tabular}{|c|c|c|c|c|}
    \hline
    1 & 2 & 4 & 5 & 8 \\
    \hline
    3 & 6 & 7 \\
    \cline{1-3}
\end{tabular},
S_{18} = 
\begin{tabular}{|c|c|c|c|c|}
    \hline
    1 & 3 & 4 & 5 & 8 \\
    \hline
    2 & 6 & 7 \\
    \cline{1-3}
\end{tabular},\\
S_{19} = 
\begin{tabular}{|c|c|c|c|c|}
    \hline
    1 & 2 & 3 & 6 & 8 \\
    \hline
    4 & 5 & 7 \\
    \cline{1-3}
\end{tabular},
S_{20} = 
\begin{tabular}{|c|c|c|c|c|}
    \hline
    1 & 2 & 4 & 6 & 8 \\
    \hline
    3 & 5 & 7 \\
    \cline{1-3}
\end{tabular},
S_{21} = 
\begin{tabular}{|c|c|c|c|c|}
    \hline
    1 & 3 & 4 & 6 & 8 \\
    \hline
    2 & 5 & 7 \\
    \cline{1-3}
\end{tabular},\\
S_{22} = 
\begin{tabular}{|c|c|c|c|c|}
    \hline
    1 & 2 & 5 & 6 & 8 \\
    \hline
    3 & 4 & 7 \\
    \cline{1-3}
\end{tabular},
S_{23} = 
\begin{tabular}{|c|c|c|c|c|}
    \hline
    1 & 3 & 5 & 6 & 8 \\
    \hline
    2 & 4 & 7 \\
    \cline{1-3}
\end{tabular},
S_{24} = 
\begin{tabular}{|c|c|c|c|c|}
    \hline
    1 & 2 & 3 & 7 & 8 \\
    \hline
    4 & 5 & 6 \\
    \cline{1-3}
\end{tabular},\\
S_{25} = 
\begin{tabular}{|c|c|c|c|c|}
    \hline
    1 & 2 & 4 & 7 & 8 \\
    \hline
    3 & 5 & 6 \\
    \cline{1-3}
\end{tabular},
S_{26} = 
\begin{tabular}{|c|c|c|c|c|}
    \hline
    1 & 3 & 4 & 7 & 8 \\
    \hline
    2 & 5 & 6 \\
    \cline{1-3}
\end{tabular},
S_{27} = 
\begin{tabular}{|c|c|c|c|c|}
    \hline
    1 & 2 & 5 & 7 & 8 \\
    \hline
    3 & 4 & 6 \\
    \cline{1-3}
\end{tabular},\\
S_{28} = 
\begin{tabular}{|c|c|c|c|c|}
    \hline
    1 & 3 & 5 & 7 & 8 \\
    \hline
    2 & 4 & 6 \\
    \cline{1-3}
\end{tabular}$.
\end{small}

\subsection{$S=0$}

\noindent $S_1 = 
\begin{tabular}{|c|c|c|c|}
    \hline
    1 & 2 & 3 & 4 \\
    \hline
    5 & 6 & 7 & 8 \\
    \hline
\end{tabular},
S_2 = 
\begin{tabular}{|c|c|c|c|}
    \hline
    1 & 2 & 3 & 5 \\
    \hline
    4 & 6 & 7 & 8 \\
    \hline
\end{tabular},
S_3 = 
\begin{tabular}{|c|c|c|c|}
    \hline
    1 & 2 & 4 & 5 \\
    \hline
    3 & 6 & 7 & 8 \\
    \hline
\end{tabular},\\
S_4 = 
\begin{tabular}{|c|c|c|c|}
    \hline
    1 & 3 & 4 & 5 \\
    \hline
    2 & 6 & 7 & 8 \\
    \hline
\end{tabular},
S_5 = 
\begin{tabular}{|c|c|c|c|}
    \hline
    1 & 2 & 3 & 6 \\
    \hline
    4 & 5 & 7 & 8 \\
    \hline
\end{tabular},
S_6 = 
\begin{tabular}{|c|c|c|c|}
    \hline
    1 & 2 & 4 & 6 \\
    \hline
    3 & 5 & 7 & 8 \\
    \hline
\end{tabular},\\
S_7 = 
\begin{tabular}{|c|c|c|c|}
    \hline
    1 & 3 & 4 & 6 \\
    \hline
    2 & 5 & 7 & 8 \\
    \hline
\end{tabular},
S_8 = 
\begin{tabular}{|c|c|c|c|}
    \hline
    1 & 2 & 5 & 6 \\
    \hline
    3 & 4 & 7 & 8 \\
    \hline
\end{tabular},
S_9 = 
\begin{tabular}{|c|c|c|c|}
    \hline
    1 & 3 & 5 & 6 \\
    \hline
    2 & 4 & 7 & 8 \\
    \hline
\end{tabular},\\
S_{10} = 
\begin{tabular}{|c|c|c|c|}
    \hline
    1 & 2 & 3 & 7 \\
    \hline
    4 & 5 & 6 & 8 \\
    \hline
\end{tabular},
S_{11} = 
\begin{tabular}{|c|c|c|c|}
    \hline
    1 & 2 & 4 & 7 \\
    \hline
    3 & 5 & 6 & 8 \\
    \hline
\end{tabular},
S_{12} = 
\begin{tabular}{|c|c|c|c|}
    \hline
    1 & 3 & 4 & 7 \\
    \hline
    2 & 5 & 6 & 8 \\
    \hline
\end{tabular},\\
S_{13} = 
\begin{tabular}{|c|c|c|c|}
    \hline
    1 & 2 & 5 & 7 \\
    \hline
    3 & 4 & 6 & 8 \\
    \hline
\end{tabular},
S_{14} = 
\begin{tabular}{|c|c|c|c|}
    \hline
    1 & 3 & 5 & 7 \\
    \hline
    2 & 4 & 6 & 8 \\
    \hline
\end{tabular}$.

\section{Wave function of octaquarks}
\label{Wave function of a octaquark}

\subsection{$q^7 \overline{Q}(I=\frac{5}{2})$}

In this section, we represent the wave function of octaquarks using Young diagrams. Here, we leave the box empty for the $u$ and $d$ light quarks, while for the strange quark and heavy quarks, we enter numbers inside the box to distinguish between them.

\begin{small}
\begin{align}
    F=\left( 
    \begin{tabular}{|c|c|c|c|c|c|}
         \hline
         \quad \quad & \quad \quad & \quad \quad & \quad \quad & \quad \quad & \quad \quad \\
         \hline
         \quad \quad  \\
         \cline{1-1}
    \end{tabular}, 
    \begin{tabular}{|c|}
         \hline
         $\hspace{0.07cm} \overline{8}\hspace{0.07cm}$  \\
         \hline
    \end{tabular}\right),\
    CS= \left(
    \begin{tabular}{|c|c|}
         \hline
         \quad \quad & \quad \quad \\
         \hline
         \quad \quad \\
         \cline{1-1}
         \quad \quad \\
         \cline{1-1}
         \quad \quad \\
         \cline{1-1}
         \quad \quad \\
         \cline{1-1}
         \quad \quad \\
         \cline{1-1}
    \end{tabular},
    \begin{tabular}{|c|}
         \hline
         $\hspace{0.07cm} \overline{8}\hspace{0.07cm}$  \\
         \hline
    \end{tabular}\right)
\end{align}
\end{small}

Now, we can determine the color and spin decomposition from color-spin coupling state.

\subsubsection{S=1}

\begin{small}
\begin{align}
    CS_1=\left(
    \begin{tabular}{|c|c|c|}
         \hline
         \quad \quad & \quad \quad & \quad \quad  \\
         \hline
         \quad \quad & \quad \quad \\
         \cline{1-2}
         \quad \quad & \quad \quad \\
         \cline{1-2}
    \end{tabular},
    \begin{tabular}{|c|}
         \hline
         $\hspace{0.07cm} \overline{8}\hspace{0.07cm}$  \\
         \hline
    \end{tabular}\right)_C \otimes
    \begin{tabular}{|c|c|c|c|c|}
         \hline
         \quad \quad & \quad \quad & \quad \quad & \quad \quad & $\hspace{0.07cm} \overline{8}\hspace{0.07cm}$   \\
         \hline
         \quad \quad & \quad \quad & \quad \quad \\
         \cline{1-3}
    \end{tabular}_S
\end{align}
\end{small}

\subsubsection{S=0}

\begin{small}
\begin{align}
    CS_1=\left(
    \begin{tabular}{|c|c|c|}
         \hline
         \quad \quad & \quad \quad & \quad \quad  \\
         \hline
         \quad \quad & \quad \quad \\
         \cline{1-2}
         \quad \quad & \quad \quad \\
         \cline{1-2}
    \end{tabular},
    \begin{tabular}{|c|}
         \hline
         $\hspace{0.07cm} \overline{8}\hspace{0.07cm}$  \\
         \hline
    \end{tabular}\right)_C \otimes
    \begin{tabular}{|c|c|c|c|}
         \hline
         \quad \quad & \quad \quad & \quad \quad & \quad \quad    \\
         \hline
         \quad \quad & \quad \quad & \quad \quad & $\hspace{0.07cm} \overline{8}\hspace{0.07cm}$ \\
         \hline
    \end{tabular}_S
\end{align}
\end{small}

\subsection{$q^7 \overline{Q} (I=\frac{3}{2})$}

\begin{small}
\begin{align}
    F=\left( 
    \begin{tabular}{|c|c|c|c|c|}
         \hline
         \quad \quad & \quad \quad & \quad \quad & \quad \quad & \quad \quad \\
         \hline
         \quad \quad & \quad \quad  \\
         \cline{1-2}
    \end{tabular}, 
    \begin{tabular}{|c|}
         \hline
         $\hspace{0.07cm} \overline{8}\hspace{0.07cm}$  \\
         \hline
    \end{tabular}\right),\
    CS= \left(
    \begin{tabular}{|c|c|}
         \hline
         \quad \quad & \quad \quad \\
         \hline
         \quad \quad & \quad \quad \\
         \cline{1-2}
         \quad \quad \\
         \cline{1-1}
         \quad \quad \\
         \cline{1-1}
         \quad \quad \\
         \cline{1-1}
    \end{tabular},
    \begin{tabular}{|c|}
         \hline
         $\hspace{0.07cm} \overline{8}\hspace{0.07cm}$  \\
         \hline
    \end{tabular}\right)
\end{align}
\end{small}

\subsubsection{$S=2$}

\begin{small}
\begin{align}
    CS_1=\left(
    \begin{tabular}{|c|c|c|}
         \hline
         \quad \quad & \quad \quad & \quad \quad  \\
         \hline
         \quad \quad & \quad \quad \\
         \cline{1-2}
         \quad \quad & \quad \quad \\
         \cline{1-2}
    \end{tabular},
    \begin{tabular}{|c|}
         \hline
         $\hspace{0.07cm} \overline{8}\hspace{0.07cm}$  \\
         \hline
    \end{tabular}\right)_C \otimes
    \begin{tabular}{|c|c|c|c|c|c|}
         \hline
         \quad \quad & \quad \quad & \quad \quad & \quad \quad & \quad \quad & $\hspace{0.07cm} \overline{8}\hspace{0.07cm}$    \\
         \hline
         \quad \quad & \quad \quad  \\
         \cline{1-2}
    \end{tabular}_S
\end{align}
\end{small}

\subsubsection{$S=1$}

\begin{small}
\begin{align}
    CS_1 &= \left(
    \begin{tabular}{|c|c|c|}
         \hline
         \quad \quad & \quad \quad & \quad \quad  \\
         \hline
         \quad \quad & \quad \quad \\
         \cline{1-2}
         \quad \quad & \quad \quad \\
         \cline{1-2}
    \end{tabular},
    \begin{tabular}{|c|}
         \hline
         $\hspace{0.07cm} \overline{8}\hspace{0.07cm}$  \\
         \hline
    \end{tabular}\right)_C \otimes
    \begin{tabular}{|c|c|c|c|c|}
         \hline
         \quad \quad & \quad \quad & \quad \quad & \quad \quad & \quad \quad     \\
         \hline
         \quad \quad & \quad \quad & $\hspace{0.07cm} \overline{8}\hspace{0.07cm}$ \\
         \cline{1-3}
    \end{tabular}_S \\
    CS_2 &= \left(
    \begin{tabular}{|c|c|c|}
         \hline
         \quad \quad & \quad \quad & \quad \quad  \\
         \hline
         \quad \quad & \quad \quad \\
         \cline{1-2}
         \quad \quad & \quad \quad \\
         \cline{1-2}
    \end{tabular},
    \begin{tabular}{|c|}
         \hline
         $\hspace{0.07cm} \overline{8}\hspace{0.07cm}$  \\
         \hline
    \end{tabular}\right)_C \otimes
    \begin{tabular}{|c|c|c|c|c|}
         \hline
         \quad \quad & \quad \quad & \quad \quad & \quad \quad & $\hspace{0.07cm} \overline{8}\hspace{0.07cm}$     \\
         \hline
         \quad \quad & \quad \quad & \quad \quad  \\
         \cline{1-3}
    \end{tabular}_S
\end{align}
\end{small}

\subsubsection{$S=0$}

\begin{small}
\begin{align}
    CS_1 &= \left(
    \begin{tabular}{|c|c|c|}
         \hline
         \quad \quad & \quad \quad & \quad \quad  \\
         \hline
         \quad \quad & \quad \quad \\
         \cline{1-2}
         \quad \quad & \quad \quad \\
         \cline{1-2}
    \end{tabular},
    \begin{tabular}{|c|}
         \hline
         $\hspace{0.07cm} \overline{8}\hspace{0.07cm}$  \\
         \hline
    \end{tabular}\right)_C \otimes
    \begin{tabular}{|c|c|c|c|}
         \hline
         \quad \quad & \quad \quad & \quad \quad & \quad \quad      \\
         \hline
         \quad \quad & \quad \quad & \quad \quad& $\hspace{0.07cm} \overline{8}\hspace{0.07cm}$ \\
         \hline
    \end{tabular}_S 
\end{align}
\end{small}

\subsection{$q^7 \overline{Q} (I=\frac{1}{2})$ }

\begin{small}
\begin{align}
    F=\left( 
    \begin{tabular}{|c|c|c|c|}
         \hline
         \quad \quad & \quad \quad & \quad \quad & \quad \quad  \\
         \hline
         \quad \quad & \quad \quad & \quad \quad  \\
         \cline{1-3}
    \end{tabular}, 
    \begin{tabular}{|c|}
         \hline
         $\hspace{0.07cm} \overline{8}\hspace{0.07cm}$  \\
         \hline
    \end{tabular}\right),\
    CS= \left(
    \begin{tabular}{|c|c|}
         \hline
         \quad \quad & \quad \quad \\
         \hline
         \quad \quad & \quad \quad \\
         \hline
         \quad \quad & \quad \quad \\
         \hline
         \quad \quad \\
         \cline{1-1}
    \end{tabular},
    \begin{tabular}{|c|}
         \hline
         $\hspace{0.07cm} \overline{8}\hspace{0.07cm}$  \\
         \hline
    \end{tabular}\right)
\end{align}
\end{small}

\subsubsection{$S=3$}

\begin{small}
\begin{align}
    CS_1 &= \left(
    \begin{tabular}{|c|c|c|}
         \hline
         \quad \quad & \quad \quad & \quad \quad  \\
         \hline
         \quad \quad & \quad \quad \\
         \cline{1-2}
         \quad \quad & \quad \quad \\
         \cline{1-2}
    \end{tabular},
    \begin{tabular}{|c|}
         \hline
         $\hspace{0.07cm} \overline{8}\hspace{0.07cm}$  \\
         \hline
    \end{tabular}\right)_C \otimes
    \begin{tabular}{|c|c|c|c|c|c|c|}
         \hline
         \quad \quad & \quad \quad & \quad \quad & \quad \quad & \quad \quad & \quad \quad & $\hspace{0.07cm} \overline{8}\hspace{0.07cm}$     \\
         \hline
         \quad \quad   \\
         \cline{1-1}
    \end{tabular}_S 
\end{align}
\end{small}

\subsubsection{$S=2$}

\begin{small}
\begin{align}
    CS_1 &= \left(
    \begin{tabular}{|c|c|c|}
         \hline
         \quad \quad & \quad \quad & \quad \quad  \\
         \hline
         \quad \quad & \quad \quad \\
         \cline{1-2}
         \quad \quad & \quad \quad \\
         \cline{1-2}
    \end{tabular},
    \begin{tabular}{|c|}
         \hline
         $\hspace{0.07cm} \overline{8}\hspace{0.07cm}$  \\
         \hline
    \end{tabular}\right)_C \otimes
    \begin{tabular}{|c|c|c|c|c|c|}
         \hline
         \quad \quad & \quad \quad & \quad \quad & \quad \quad & \quad \quad & \quad \quad      \\
         \hline
         \quad \quad & $\hspace{0.07cm} \overline{8}\hspace{0.07cm}$  \\
         \cline{1-2}
    \end{tabular}_S \\
    CS_2 &= \left(
    \begin{tabular}{|c|c|c|}
         \hline
         \quad \quad & \quad \quad & \quad \quad  \\
         \hline
         \quad \quad & \quad \quad \\
         \cline{1-2}
         \quad \quad & \quad \quad \\
         \cline{1-2}
    \end{tabular},
    \begin{tabular}{|c|}
         \hline
         $\hspace{0.07cm} \overline{8}\hspace{0.07cm}$  \\
         \hline
    \end{tabular}\right)_C \otimes
    \begin{tabular}{|c|c|c|c|c|c|}
         \hline
         \quad \quad & \quad \quad & \quad \quad & \quad \quad & \quad \quad & $\hspace{0.07cm} \overline{8}\hspace{0.07cm}$      \\
         \hline
         \quad \quad & \quad \quad  \\
         \cline{1-2}
    \end{tabular}_S 
\end{align}
\end{small}

\subsubsection{$S=1$}

\begin{small}
\begin{align}
    CS_1 &= \left(
    \begin{tabular}{|c|c|c|}
         \hline
         \quad \quad & \quad \quad & \quad \quad  \\
         \hline
         \quad \quad & \quad \quad \\
         \cline{1-2}
         \quad \quad & \quad \quad \\
         \cline{1-2}
    \end{tabular},
    \begin{tabular}{|c|}
         \hline
         $\hspace{0.07cm} \overline{8}\hspace{0.07cm}$  \\
         \hline
    \end{tabular}\right)_C \otimes
    \begin{tabular}{|c|c|c|c|c|}
         \hline
         \quad \quad & \quad \quad & \quad \quad & \quad \quad & \quad \quad      \\
         \hline
         \quad \quad & \quad \quad & $\hspace{0.07cm} \overline{8}\hspace{0.07cm}$  \\
         \cline{1-3}
    \end{tabular}_S \\
    CS_2 &= \left(
    \begin{tabular}{|c|c|c|}
         \hline
         \quad \quad & \quad \quad & \quad \quad  \\
         \hline
         \quad \quad & \quad \quad \\
         \cline{1-2}
         \quad \quad & \quad \quad \\
         \cline{1-2}
    \end{tabular},
    \begin{tabular}{|c|}
         \hline
         $\hspace{0.07cm} \overline{8}\hspace{0.07cm}$  \\
         \hline
    \end{tabular}\right)_C \otimes
    \begin{tabular}{|c|c|c|c|c|}
         \hline
         \quad \quad & \quad \quad & \quad \quad & \quad \quad & $\hspace{0.07cm} \overline{8}\hspace{0.07cm}$      \\
         \hline
         \quad \quad & \quad \quad & \quad \quad  \\
         \cline{1-3}
    \end{tabular}_S 
\end{align}
\end{small}

\subsubsection{$S=0$}

\begin{small}
\begin{align}
    CS_1 &= \left(
    \begin{tabular}{|c|c|c|}
         \hline
         \quad \quad & \quad \quad & \quad \quad  \\
         \hline
         \quad \quad & \quad \quad \\
         \cline{1-2}
         \quad \quad & \quad \quad \\
         \cline{1-2}
    \end{tabular},
    \begin{tabular}{|c|}
         \hline
         $\hspace{0.07cm} \overline{8}\hspace{0.07cm}$  \\
         \hline
    \end{tabular}\right)_C \otimes
    \begin{tabular}{|c|c|c|c|}
         \hline
         \quad \quad & \quad \quad & \quad \quad & \quad \quad \\
         \hline
         \quad \quad & \quad \quad & \quad \quad & $\hspace{0.07cm} \overline{8}\hspace{0.07cm}$  \\
         \hline
    \end{tabular}_S 
\end{align}
\end{small}

\subsection{$q^6 s \overline{Q}(I=3)$}

Here we consider the octaquark containing one strange quark. We fix the position of a strange quark to 7. Depending on where position 7 of the strange quark is located among the 21 color basis functions, the color basis set can be divided into two parts. As the number of strange quarks increases, the number of possible subsets of color bases  increases.

\begin{small}
\begin{align}
    \begin{tabular}{|c|c|c|}
         \hline
         \quad \quad & \quad \quad & \quad \quad  \\
         \hline
         \quad \quad & \quad \quad \\
         \cline{1-2}
         \quad \quad & $\hspace{0.07cm} 7\hspace{0.07cm}$ \\
         \cline{1-2}
    \end{tabular}: C_{1} - C_{16}, \quad
    \begin{tabular}{|c|c|c|}
         \hline
         \quad \quad & \quad \quad & $\hspace{0.07cm} 7\hspace{0.07cm}$  \\
         \hline
         \quad \quad & \quad \quad \\
         \cline{1-2}
         \quad \quad & \quad \quad \\
         \cline{1-2}
    \end{tabular}: C_{17} - C_{21}
\end{align}

\begin{align}
    F=\left( 
    \begin{tabular}{|c|c|c|c|c|c|}
         \hline
         \quad \quad & \quad \quad & \quad \quad & \quad \quad & \quad \quad & \quad \quad \\
         \hline
    \end{tabular}, 
    \begin{tabular}{|c|}
         \hline
         $\hspace{0.07cm} 7\hspace{0.07cm}$  \\
         \hline
    \end{tabular},
    \begin{tabular}{|c|}
         \hline
         $\hspace{0.07cm} \overline{8}\hspace{0.07cm}$  \\
         \hline
    \end{tabular}\right),\
    CS= \left(
    \begin{tabular}{|c|}
         \hline
         \quad \quad \\
         \hline
         \quad \quad \\
         \hline
         \quad \quad \\
         \hline
         \quad \quad \\
         \hline
         \quad \quad \\
         \hline
         \quad \quad \\
         \hline
    \end{tabular},
    \begin{tabular}{|c|}
         \hline
         $\hspace{0.07cm} 7\hspace{0.07cm}$  \\
         \hline
    \end{tabular},
    \begin{tabular}{|c|}
         \hline
         $\hspace{0.07cm} \overline{8}\hspace{0.07cm}$  \\
         \hline
    \end{tabular}\right)
\end{align}
\end{small}

\subsubsection{S=1}

\begin{small}
\begin{align}
    CS_1 &= \left(
    \begin{tabular}{|c|c|c|}
         \hline
         \quad \quad & \quad \quad & $\hspace{0.07cm} 7\hspace{0.07cm}$  \\
         \hline
         \quad \quad & \quad \quad \\
         \cline{1-2}
         \quad \quad & \quad \quad \\
         \cline{1-2}
    \end{tabular},
    \begin{tabular}{|c|}
         \hline
         $\hspace{0.07cm} \overline{8}\hspace{0.07cm}$  \\
         \hline
    \end{tabular}\right)_C \otimes
    \begin{tabular}{|c|c|c|c|c|}
         \hline
         \quad \quad & \quad \quad & \quad \quad & $\hspace{0.07cm} 7\hspace{0.07cm}$ & $\hspace{0.07cm} \overline{8}\hspace{0.07cm}$   \\
         \hline
         \quad \quad & \quad \quad & \quad \quad \\
         \cline{1-3}
    \end{tabular}_S
\end{align}
\end{small}

\subsubsection{S=0}

\begin{small}
\begin{align}
    CS_1 &= \left(
    \begin{tabular}{|c|c|c|}
         \hline
         \quad \quad & \quad \quad & $\hspace{0.07cm} 7\hspace{0.07cm}$  \\
         \hline
         \quad \quad & \quad \quad \\
         \cline{1-2}
         \quad \quad & \quad \quad \\
         \cline{1-2}
    \end{tabular},
    \begin{tabular}{|c|}
         \hline
         $\hspace{0.07cm} \overline{8}\hspace{0.07cm}$  \\
         \hline
    \end{tabular}\right)_C \otimes
    \begin{tabular}{|c|c|c|c|}
         \hline
         \quad \quad & \quad \quad & \quad \quad & $\hspace{0.07cm} 7\hspace{0.07cm}$    \\
         \hline
         \quad \quad & \quad \quad & \quad \quad & $\hspace{0.07cm} \overline{8}\hspace{0.07cm}$ \\
         \hline
    \end{tabular}_S
\end{align}
\end{small}

\subsection{$q^6 s \overline{Q}(I=2)$}

\begin{small}
\begin{align}
    F=\left( 
    \begin{tabular}{|c|c|c|c|c|}
         \hline
         \quad \quad & \quad \quad & \quad \quad & \quad \quad & \quad \quad \\
         \hline
         \quad \quad \\
         \cline{1-1}
    \end{tabular}, 
    \begin{tabular}{|c|}
         \hline
         $\hspace{0.07cm} 7\hspace{0.07cm}$  \\
         \hline
    \end{tabular},
    \begin{tabular}{|c|}
         \hline
         $\hspace{0.07cm} \overline{8}\hspace{0.07cm}$  \\
         \hline
    \end{tabular}\right),\
    CS= \left(
    \begin{tabular}{|c|c|}
         \hline
         \quad \quad & \quad \quad \\
         \hline
         \quad \quad \\
         \cline{1-1}
         \quad \quad \\
         \cline{1-1}
         \quad \quad \\
         \cline{1-1}
         \quad \quad \\
         \cline{1-1}
    \end{tabular},
    \begin{tabular}{|c|}
         \hline
         $\hspace{0.07cm} 7\hspace{0.07cm}$  \\
         \hline
    \end{tabular},
    \begin{tabular}{|c|}
         \hline
         $\hspace{0.07cm} \overline{8}\hspace{0.07cm}$  \\
         \hline
    \end{tabular}\right)
\end{align}
\end{small}

\subsubsection{S=2}

\begin{small}
\begin{align}
    CS_1 &= \left(
    \begin{tabular}{|c|c|c|}
         \hline
         \quad \quad & \quad \quad & \quad \quad  \\
         \hline
         \quad \quad & \quad \quad \\
         \cline{1-2}
         \quad \quad & $\hspace{0.07cm} 7\hspace{0.07cm}$ \\
         \cline{1-2}
    \end{tabular},
    \begin{tabular}{|c|}
         \hline
         $\hspace{0.07cm} \overline{8}\hspace{0.07cm}$  \\
         \hline
    \end{tabular}\right)_C \otimes
    \begin{tabular}{|c|c|c|c|c|c|}
         \hline
         \quad \quad & \quad \quad & \quad \quad & \quad \quad & $\hspace{0.07cm} 7\hspace{0.07cm}$ & $\hspace{0.07cm} \overline{8}\hspace{0.07cm}$   \\
         \hline
         \quad \quad & \quad \quad \\
         \cline{1-2}
    \end{tabular}_S \\
    CS_2 &= \left(
    \begin{tabular}{|c|c|c|}
         \hline
         \quad \quad & \quad \quad & $\hspace{0.07cm} 7\hspace{0.07cm}$  \\
         \hline
         \quad \quad & \quad \quad \\
         \cline{1-2}
         \quad \quad & \quad \quad \\
         \cline{1-2}
    \end{tabular},
    \begin{tabular}{|c|}
         \hline
         $\hspace{0.07cm} \overline{8}\hspace{0.07cm}$  \\
         \hline
    \end{tabular}\right)_C \otimes
    \begin{tabular}{|c|c|c|c|c|c|}
         \hline
         \quad \quad & \quad \quad & \quad \quad & \quad \quad & $\hspace{0.07cm} 7\hspace{0.07cm}$ & $\hspace{0.07cm} \overline{8}\hspace{0.07cm}$   \\
         \hline
         \quad \quad & \quad \quad \\
         \cline{1-2}
    \end{tabular}_S
\end{align}
\end{small}

\subsubsection{S=1}

\begin{small}
\begin{align}
    CS_1 &= \left(
    \begin{tabular}{|c|c|c|}
         \hline
         \quad \quad & \quad \quad & \quad \quad  \\
         \hline
         \quad \quad & \quad \quad \\
         \cline{1-2}
         \quad \quad & $\hspace{0.07cm} 7\hspace{0.07cm}$ \\
         \cline{1-2}
    \end{tabular},
    \begin{tabular}{|c|}
         \hline
         $\hspace{0.07cm} \overline{8}\hspace{0.07cm}$  \\
         \hline
    \end{tabular}\right)_C \otimes
    \begin{tabular}{|c|c|c|c|c|}
         \hline
         \quad \quad & \quad \quad & \quad \quad & \quad \quad & $\hspace{0.07cm} 7\hspace{0.07cm}$    \\
         \hline
         \quad \quad & \quad \quad & $\hspace{0.07cm} \overline{8}\hspace{0.07cm}$\\
         \cline{1-3}
    \end{tabular}_S \\
    CS_2 &= \left(
    \begin{tabular}{|c|c|c|}
         \hline
         \quad \quad & \quad \quad & \quad \quad  \\
         \hline
         \quad \quad & \quad \quad \\
         \cline{1-2}
         \quad \quad & $\hspace{0.07cm} 7\hspace{0.07cm}$ \\
         \cline{1-2}
    \end{tabular},
    \begin{tabular}{|c|}
         \hline
         $\hspace{0.07cm} \overline{8}\hspace{0.07cm}$  \\
         \hline
    \end{tabular}\right)_C \otimes
    \begin{tabular}{|c|c|c|c|c|}
         \hline
         \quad \quad & \quad \quad & \quad \quad & \quad \quad & $\hspace{0.07cm} \overline{8}\hspace{0.07cm}$    \\
         \hline
         \quad \quad & \quad \quad & $\hspace{0.07cm} 7\hspace{0.07cm}$\\
         \cline{1-3}
    \end{tabular}_S \\
    CS_3 &= \left(
    \begin{tabular}{|c|c|c|}
         \hline
         \quad \quad & \quad \quad & \quad \quad  \\
         \hline
         \quad \quad & \quad \quad \\
         \cline{1-2}
         \quad \quad & $\hspace{0.07cm} 7\hspace{0.07cm}$ \\
         \cline{1-2}
    \end{tabular},
    \begin{tabular}{|c|}
         \hline
         $\hspace{0.07cm} \overline{8}\hspace{0.07cm}$  \\
         \hline
    \end{tabular}\right)_C \otimes
    \begin{tabular}{|c|c|c|c|c|}
         \hline
         \quad \quad & \quad \quad & \quad \quad & $\hspace{0.07cm} 7\hspace{0.07cm}$ & $\hspace{0.07cm} \overline{8}\hspace{0.07cm}$    \\
         \hline
         \quad \quad & \quad \quad & \quad \quad \\
         \cline{1-3}
    \end{tabular}_S \\
    CS_4 &= \left(
    \begin{tabular}{|c|c|c|}
         \hline
         \quad \quad & \quad \quad & $\hspace{0.07cm} 7\hspace{0.07cm}$  \\
         \hline
         \quad \quad & \quad \quad \\
         \cline{1-2}
         \quad \quad & \quad \quad \\
         \cline{1-2}
    \end{tabular},
    \begin{tabular}{|c|}
         \hline
         $\hspace{0.07cm} \overline{8}\hspace{0.07cm}$  \\
         \hline
    \end{tabular}\right)_C \otimes
    \begin{tabular}{|c|c|c|c|c|}
         \hline
         \quad \quad & \quad \quad & \quad \quad & \quad \quad & $\hspace{0.07cm} 7\hspace{0.07cm}$    \\
         \hline
         \quad \quad & \quad \quad & $\hspace{0.07cm} \overline{8}\hspace{0.07cm}$\\
         \cline{1-3}
    \end{tabular}_S \\
    CS_5 &= \left(
    \begin{tabular}{|c|c|c|}
         \hline
         \quad \quad & \quad \quad & $\hspace{0.07cm} 7\hspace{0.07cm}$  \\
         \hline
         \quad \quad & \quad \quad \\
         \cline{1-2}
         \quad \quad & \quad \quad \\
         \cline{1-2}
    \end{tabular},
    \begin{tabular}{|c|}
         \hline
         $\hspace{0.07cm} \overline{8}\hspace{0.07cm}$  \\
         \hline
    \end{tabular}\right)_C \otimes
    \begin{tabular}{|c|c|c|c|c|}
         \hline
         \quad \quad & \quad \quad & \quad \quad & \quad \quad & $\hspace{0.07cm} \overline{8}\hspace{0.07cm}$    \\
         \hline
         \quad \quad & \quad \quad & $\hspace{0.07cm} 7\hspace{0.07cm}$\\
         \cline{1-3}
    \end{tabular}_S 
\end{align}
\end{small}

\subsubsection{S=0}

\begin{small}
\begin{align}
    CS_1 &= \left(
    \begin{tabular}{|c|c|c|}
         \hline
         \quad \quad & \quad \quad & \quad \quad  \\
         \hline
         \quad \quad & \quad \quad \\
         \cline{1-2}
         \quad \quad & $\hspace{0.07cm} 7\hspace{0.07cm}$ \\
         \cline{1-2}
    \end{tabular},
    \begin{tabular}{|c|}
         \hline
         $\hspace{0.07cm} \overline{8}\hspace{0.07cm}$  \\
         \hline
    \end{tabular}\right)_C \otimes
    \begin{tabular}{|c|c|c|c|}
         \hline
         \quad \quad & \quad \quad & \quad \quad & \quad \quad     \\
         \hline
         \quad \quad & \quad \quad & $\hspace{0.07cm} 7\hspace{0.07cm}$ & $\hspace{0.07cm} \overline{8}\hspace{0.07cm}$\\
         \hline
    \end{tabular}_S \\
    CS_2 &= \left(
    \begin{tabular}{|c|c|c|}
         \hline
         \quad \quad & \quad \quad & \quad \quad  \\
         \hline
         \quad \quad & \quad \quad \\
         \cline{1-2}
         \quad \quad & $\hspace{0.07cm} 7\hspace{0.07cm}$ \\
         \cline{1-2}
    \end{tabular},
    \begin{tabular}{|c|}
         \hline
         $\hspace{0.07cm} \overline{8}\hspace{0.07cm}$  \\
         \hline
    \end{tabular}\right)_C \otimes
    \begin{tabular}{|c|c|c|c|}
         \hline
         \quad \quad & \quad \quad & \quad \quad & $\hspace{0.07cm} 7\hspace{0.07cm}$     \\
         \hline
         \quad \quad & \quad \quad & \quad \quad & $\hspace{0.07cm} \overline{8}\hspace{0.07cm}$\\
         \hline
    \end{tabular}_S \\ 
    CS_3 &= \left(
    \begin{tabular}{|c|c|c|}
         \hline
         \quad \quad & \quad \quad & $\hspace{0.07cm} 7\hspace{0.07cm}$  \\
         \hline
         \quad \quad & \quad \quad \\
         \cline{1-2}
         \quad \quad & \quad \quad \\
         \cline{1-2}
    \end{tabular},
    \begin{tabular}{|c|}
         \hline
         $\hspace{0.07cm} \overline{8}\hspace{0.07cm}$  \\
         \hline
    \end{tabular}\right)_C \otimes
    \begin{tabular}{|c|c|c|c|}
         \hline
         \quad \quad & \quad \quad & \quad \quad & \quad \quad     \\
         \hline
         \quad \quad & \quad \quad & $\hspace{0.07cm} 7\hspace{0.07cm}$ & $\hspace{0.07cm} \overline{8}\hspace{0.07cm}$\\
         \hline
    \end{tabular}_S 
\end{align}
\end{small}

\subsection{$q^6 s \overline{Q}(I=1)$}

\begin{small}
\begin{align}
    F=\left( 
    \begin{tabular}{|c|c|c|c|}
         \hline
         \quad \quad & \quad \quad & \quad \quad & \quad \quad  \\
         \hline
         \quad \quad & \quad \quad \\
         \cline{1-2}
    \end{tabular}, 
    \begin{tabular}{|c|}
         \hline
         $\hspace{0.07cm} 7\hspace{0.07cm}$  \\
         \hline
    \end{tabular},
    \begin{tabular}{|c|}
         \hline
         $\hspace{0.07cm} \overline{8}\hspace{0.07cm}$  \\
         \hline
    \end{tabular}\right),\
    CS= \left(
    \begin{tabular}{|c|c|}
         \hline
         \quad \quad & \quad \quad \\
         \hline
         \quad \quad & \quad \quad \\
         \hline
         \quad \quad \\
         \cline{1-1}
         \quad \quad \\
         \cline{1-1}
    \end{tabular},
    \begin{tabular}{|c|}
         \hline
         $\hspace{0.07cm} 7\hspace{0.07cm}$  \\
         \hline
    \end{tabular},
    \begin{tabular}{|c|}
         \hline
         $\hspace{0.07cm} \overline{8}\hspace{0.07cm}$  \\
         \hline
    \end{tabular}\right)
\end{align}
\end{small}

\subsubsection{S=3}

\begin{small}
\begin{align}
    CS_1 &= \left(
    \begin{tabular}{|c|c|c|}
         \hline
         \quad \quad & \quad \quad & \quad \quad  \\
         \hline
         \quad \quad & \quad \quad \\
         \cline{1-2}
         \quad \quad & $\hspace{0.07cm} 7\hspace{0.07cm}$ \\
         \cline{1-2}
    \end{tabular},
    \begin{tabular}{|c|}
         \hline
         $\hspace{0.07cm} \overline{8}\hspace{0.07cm}$  \\
         \hline
    \end{tabular}\right)_C \otimes
    \begin{tabular}{|c|c|c|c|c|c|c|}
         \hline
         \quad \quad & \quad \quad & \quad \quad & \quad \quad & \quad \quad & $\hspace{0.07cm} 7\hspace{0.07cm}$ & $\hspace{0.07cm} \overline{8}\hspace{0.07cm}$   \\
         \hline
         \quad \quad \\
         \cline{1-1}
    \end{tabular}_S \\
    CS_2 &= \left(
    \begin{tabular}{|c|c|c|}
         \hline
         \quad \quad & \quad \quad & $\hspace{0.07cm} 7\hspace{0.07cm}$  \\
         \hline
         \quad \quad & \quad \quad \\
         \cline{1-2}
         \quad \quad & \quad \quad \\
         \cline{1-2}
    \end{tabular},
    \begin{tabular}{|c|}
         \hline
         $\hspace{0.07cm} \overline{8}\hspace{0.07cm}$  \\
         \hline
    \end{tabular}\right)_C \otimes
    \begin{tabular}{|c|c|c|c|c|c|c|}
         \hline
         \quad \quad & \quad \quad & \quad \quad & \quad \quad & \quad \quad & $\hspace{0.07cm} 7\hspace{0.07cm}$ & $\hspace{0.07cm} \overline{8}\hspace{0.07cm}$   \\
         \hline
         \quad \quad \\
         \cline{1-1}
    \end{tabular}_S
\end{align}
\end{small}

\subsubsection{S=2}

\begin{small}
\begin{align}
    CS_1 &= \left(
    \begin{tabular}{|c|c|c|}
         \hline
         \quad \quad & \quad \quad & \quad \quad  \\
         \hline
         \quad \quad & \quad \quad \\
         \cline{1-2}
         \quad \quad & $\hspace{0.07cm} 7\hspace{0.07cm}$ \\
         \cline{1-2}
    \end{tabular},
    \begin{tabular}{|c|}
         \hline
         $\hspace{0.07cm} \overline{8}\hspace{0.07cm}$  \\
         \hline
    \end{tabular}\right)_C \otimes
    \begin{tabular}{|c|c|c|c|c|c|}
         \hline
         \quad \quad & \quad \quad & \quad \quad & \quad \quad & \quad \quad & $\hspace{0.07cm} 7\hspace{0.07cm}$   \\
         \hline
         \quad \quad & $\hspace{0.07cm} \overline{8}\hspace{0.07cm}$ \\
         \cline{1-2}
    \end{tabular}_S \\
    CS_2 &= \left(
    \begin{tabular}{|c|c|c|}
         \hline
         \quad \quad & \quad \quad & \quad \quad  \\
         \hline
         \quad \quad & \quad \quad \\
         \cline{1-2}
         \quad \quad & $\hspace{0.07cm} 7\hspace{0.07cm}$ \\
         \cline{1-2}
    \end{tabular},
    \begin{tabular}{|c|}
         \hline
         $\hspace{0.07cm} \overline{8}\hspace{0.07cm}$  \\
         \hline
    \end{tabular}\right)_C \otimes
    \begin{tabular}{|c|c|c|c|c|c|}
         \hline
         \quad \quad & \quad \quad & \quad \quad & \quad \quad & \quad \quad & $\hspace{0.07cm} \overline{8}\hspace{0.07cm}$   \\
         \hline
         \quad \quad & $\hspace{0.07cm} 7\hspace{0.07cm}$ \\
         \cline{1-2}
    \end{tabular}_S \\
    CS_{3,4} &= \left(
    \begin{tabular}{|c|c|c|}
         \hline
         \quad \quad & \quad \quad & \quad \quad  \\
         \hline
         \quad \quad & \quad \quad \\
         \cline{1-2}
         \quad \quad & $\hspace{0.07cm} 7\hspace{0.07cm}$ \\
         \cline{1-2}
    \end{tabular},
    \begin{tabular}{|c|}
         \hline
         $\hspace{0.07cm} \overline{8}\hspace{0.07cm}$  \\
         \hline
    \end{tabular}\right)_C \otimes
    \begin{tabular}{|c|c|c|c|c|c|}
         \hline
         \quad \quad & \quad \quad & \quad \quad & \quad \quad & $\hspace{0.07cm} 7\hspace{0.07cm}$ & $\hspace{0.07cm} \overline{8}\hspace{0.07cm}$   \\
         \hline
         \quad \quad & \quad \quad \\
         \cline{1-2}
    \end{tabular}_S \quad (m=2)\\
    CS_5 &= \left(
    \begin{tabular}{|c|c|c|}
         \hline
         \quad \quad & \quad \quad & $\hspace{0.07cm} 7\hspace{0.07cm}$  \\
         \hline
         \quad \quad & \quad \quad \\
         \cline{1-2}
         \quad \quad & \quad \quad \\
         \cline{1-2}
    \end{tabular},
    \begin{tabular}{|c|}
         \hline
         $\hspace{0.07cm} \overline{8}\hspace{0.07cm}$  \\
         \hline
    \end{tabular}\right)_C \otimes
    \begin{tabular}{|c|c|c|c|c|c|}
         \hline
         \quad \quad & \quad \quad & \quad \quad & \quad \quad & \quad \quad & $\hspace{0.07cm} 7\hspace{0.07cm}$   \\
         \hline
         \quad \quad & $\hspace{0.07cm} \overline{8}\hspace{0.07cm}$ \\
         \cline{1-2}
    \end{tabular}_S \\
    CS_6 &= \left(
    \begin{tabular}{|c|c|c|}
         \hline
         \quad \quad & \quad \quad & $\hspace{0.07cm} 7\hspace{0.07cm}$  \\
         \hline
         \quad \quad & \quad \quad \\
         \cline{1-2}
         \quad \quad & \quad \quad \\
         \cline{1-2}
    \end{tabular},
    \begin{tabular}{|c|}
         \hline
         $\hspace{0.07cm} \overline{8}\hspace{0.07cm}$  \\
         \hline
    \end{tabular}\right)_C \otimes
    \begin{tabular}{|c|c|c|c|c|c|}
         \hline
         \quad \quad & \quad \quad & \quad \quad & \quad \quad & \quad \quad & $\hspace{0.07cm} \overline{8}\hspace{0.07cm}$   \\
         \hline
         \quad \quad & $\hspace{0.07cm} 7\hspace{0.07cm}$ \\
         \cline{1-2}
    \end{tabular}_S 
\end{align}
\end{small}

\subsubsection{S=1}

\begin{small}
\begin{align}
    CS_1 &= \left(
    \begin{tabular}{|c|c|c|}
         \hline
         \quad \quad & \quad \quad & \quad \quad  \\
         \hline
         \quad \quad & \quad \quad \\
         \cline{1-2}
         \quad \quad & $\hspace{0.07cm} 7\hspace{0.07cm}$ \\
         \cline{1-2}
    \end{tabular},
    \begin{tabular}{|c|}
         \hline
         $\hspace{0.07cm} \overline{8}\hspace{0.07cm}$  \\
         \hline
    \end{tabular}\right)_C \otimes
    \begin{tabular}{|c|c|c|c|c|}
         \hline
         \quad \quad & \quad \quad & \quad \quad & \quad \quad & \quad \quad    \\
         \hline
         \quad \quad & $\hspace{0.07cm} 7\hspace{0.07cm}$ & $\hspace{0.07cm} \overline{8}\hspace{0.07cm}$\\
         \cline{1-3}
    \end{tabular}_S \\
    CS_{2,3} &= \left(
    \begin{tabular}{|c|c|c|}
         \hline
         \quad \quad & \quad \quad & \quad \quad  \\
         \hline
         \quad \quad & \quad \quad \\
         \cline{1-2}
         \quad \quad & $\hspace{0.07cm} 7\hspace{0.07cm}$ \\
         \cline{1-2}
    \end{tabular},
    \begin{tabular}{|c|}
         \hline
         $\hspace{0.07cm} \overline{8}\hspace{0.07cm}$  \\
         \hline
    \end{tabular}\right)_C \otimes
    \begin{tabular}{|c|c|c|c|c|}
         \hline
         \quad \quad & \quad \quad & \quad \quad & \quad \quad & $\hspace{0.07cm} 7\hspace{0.07cm}$    \\
         \hline
         \quad \quad & \quad \quad & $\hspace{0.07cm} \overline{8}\hspace{0.07cm}$\\
         \cline{1-3}
    \end{tabular}_S \quad (m=2) \\
    CS_{4,5} &= \left(
    \begin{tabular}{|c|c|c|}
         \hline
         \quad \quad & \quad \quad & \quad \quad  \\
         \hline
         \quad \quad & \quad \quad \\
         \cline{1-2}
         \quad \quad & $\hspace{0.07cm} 7\hspace{0.07cm}$ \\
         \cline{1-2}
    \end{tabular},
    \begin{tabular}{|c|}
         \hline
         $\hspace{0.07cm} \overline{8}\hspace{0.07cm}$  \\
         \hline
    \end{tabular}\right)_C \otimes
    \begin{tabular}{|c|c|c|c|c|}
         \hline
         \quad \quad & \quad \quad & \quad \quad & \quad \quad & $\hspace{0.07cm} \overline{8}\hspace{0.07cm}$    \\
         \hline
         \quad \quad & \quad \quad & $\hspace{0.07cm} 7\hspace{0.07cm}$\\
         \cline{1-3}
    \end{tabular}_S \quad (m=2) \\
    CS_6 &= \left(
    \begin{tabular}{|c|c|c|}
         \hline
         \quad \quad & \quad \quad & \quad \quad  \\
         \hline
         \quad \quad & \quad \quad \\
         \cline{1-2}
         \quad \quad & $\hspace{0.07cm} 7\hspace{0.07cm}$ \\
         \cline{1-2}
    \end{tabular},
    \begin{tabular}{|c|}
         \hline
         $\hspace{0.07cm} \overline{8}\hspace{0.07cm}$  \\
         \hline
    \end{tabular}\right)_C \otimes
    \begin{tabular}{|c|c|c|c|c|}
         \hline
         \quad \quad & \quad \quad & \quad \quad & $\hspace{0.07cm} 7\hspace{0.07cm}$ & $\hspace{0.07cm} \overline{8}\hspace{0.07cm}$    \\
         \hline
         \quad \quad & \quad \quad & \quad \quad \\
         \cline{1-3}
    \end{tabular}_S \\
    CS_7 &= \left(
    \begin{tabular}{|c|c|c|}
         \hline
         \quad \quad & \quad \quad & $\hspace{0.07cm} 7\hspace{0.07cm}$  \\
         \hline
         \quad \quad & \quad \quad \\
         \cline{1-2}
         \quad \quad & \quad \quad \\
         \cline{1-2}
    \end{tabular},
    \begin{tabular}{|c|}
         \hline
         $\hspace{0.07cm} \overline{8}\hspace{0.07cm}$  \\
         \hline
    \end{tabular}\right)_C \otimes
    \begin{tabular}{|c|c|c|c|c|}
         \hline
         \quad \quad & \quad \quad & \quad \quad & \quad \quad & \quad \quad    \\
         \hline
         \quad \quad & $\hspace{0.07cm} 7\hspace{0.07cm}$ & $\hspace{0.07cm} \overline{8}\hspace{0.07cm}$\\
         \cline{1-3}
    \end{tabular}_S \\
    CS_8 &= \left(
    \begin{tabular}{|c|c|c|}
         \hline
         \quad \quad & \quad \quad & $\hspace{0.07cm} 7\hspace{0.07cm}$  \\
         \hline
         \quad \quad & \quad \quad \\
         \cline{1-2}
         \quad \quad & \quad \quad \\
         \cline{1-2}
    \end{tabular},
    \begin{tabular}{|c|}
         \hline
         $\hspace{0.07cm} \overline{8}\hspace{0.07cm}$  \\
         \hline
    \end{tabular}\right)_C \otimes
    \begin{tabular}{|c|c|c|c|c|}
         \hline
         \quad \quad & \quad \quad & \quad \quad & $\hspace{0.07cm} 7\hspace{0.07cm}$ & $\hspace{0.07cm} \overline{8}\hspace{0.07cm}$    \\
         \hline
         \quad \quad & \quad \quad & \quad \quad \\
         \cline{1-3}
    \end{tabular}_S 
\end{align}
\end{small}

\subsubsection{S=0}

\begin{small}
\begin{align}
    CS_{1,2} &= \left(
    \begin{tabular}{|c|c|c|}
         \hline
         \quad \quad & \quad \quad & \quad \quad  \\
         \hline
         \quad \quad & \quad \quad \\
         \cline{1-2}
         \quad \quad & $\hspace{0.07cm} 7\hspace{0.07cm}$ \\
         \cline{1-2}
    \end{tabular},
    \begin{tabular}{|c|}
         \hline
         $\hspace{0.07cm} \overline{8}\hspace{0.07cm}$  \\
         \hline
    \end{tabular}\right)_C \otimes
    \begin{tabular}{|c|c|c|c|}
         \hline
         \quad \quad & \quad \quad & \quad \quad & \quad \quad     \\
         \hline
         \quad \quad & \quad \quad & $\hspace{0.07cm} 7\hspace{0.07cm}$ & $\hspace{0.07cm} \overline{8}\hspace{0.07cm}$\\
         \hline
    \end{tabular}_S \quad (m=2)\\
    CS_3 &= \left(
    \begin{tabular}{|c|c|c|}
         \hline
         \quad \quad & \quad \quad & \quad \quad  \\
         \hline
         \quad \quad & \quad \quad \\
         \cline{1-2}
         \quad \quad & $\hspace{0.07cm} 7\hspace{0.07cm}$ \\
         \cline{1-2}
    \end{tabular},
    \begin{tabular}{|c|}
         \hline
         $\hspace{0.07cm} \overline{8}\hspace{0.07cm}$  \\
         \hline
    \end{tabular}\right)_C \otimes
    \begin{tabular}{|c|c|c|c|}
         \hline
         \quad \quad & \quad \quad & \quad \quad & $\hspace{0.07cm} 7\hspace{0.07cm}$     \\
         \hline
         \quad \quad & \quad \quad & \quad \quad & $\hspace{0.07cm} \overline{8}\hspace{0.07cm}$\\
         \hline
    \end{tabular}_S \\ 
    CS_4 &= \left(
    \begin{tabular}{|c|c|c|}
         \hline
         \quad \quad & \quad \quad & $\hspace{0.07cm} 7\hspace{0.07cm}$  \\
         \hline
         \quad \quad & \quad \quad \\
         \cline{1-2}
         \quad \quad & \quad \quad \\
         \cline{1-2}
    \end{tabular},
    \begin{tabular}{|c|}
         \hline
         $\hspace{0.07cm} \overline{8}\hspace{0.07cm}$  \\
         \hline
    \end{tabular}\right)_C \otimes
    \begin{tabular}{|c|c|c|c|}
         \hline
         \quad \quad & \quad \quad & \quad \quad & $\hspace{0.07cm} 7\hspace{0.07cm}$     \\
         \hline
         \quad \quad & \quad \quad & \quad \quad & $\hspace{0.07cm} \overline{8}\hspace{0.07cm}$\\
         \hline
    \end{tabular}_S 
\end{align}
\end{small}

\subsection{$q^6 s \overline{Q}(I=0)$}

\begin{small}
\begin{align}
    F=\left( 
    \begin{tabular}{|c|c|c|}
         \hline
         \quad \quad & \quad \quad & \quad \quad   \\
         \hline
         \quad \quad & \quad \quad & \quad \quad \\
         \hline
    \end{tabular}, 
    \begin{tabular}{|c|}
         \hline
         $\hspace{0.07cm} 7\hspace{0.07cm}$  \\
         \hline
    \end{tabular},
    \begin{tabular}{|c|}
         \hline
         $\hspace{0.07cm} \overline{8}\hspace{0.07cm}$  \\
         \hline
    \end{tabular}\right),\
    CS= \left(
    \begin{tabular}{|c|c|}
         \hline
         \quad \quad & \quad \quad \\
         \hline
         \quad \quad & \quad \quad \\
         \hline
         \quad \quad & \quad \quad \\
         \hline
    \end{tabular},
    \begin{tabular}{|c|}
         \hline
         $\hspace{0.07cm} 7\hspace{0.07cm}$  \\
         \hline
    \end{tabular},
    \begin{tabular}{|c|}
         \hline
         $\hspace{0.07cm} \overline{8}\hspace{0.07cm}$  \\
         \hline
    \end{tabular}\right)
\end{align}
\end{small}

\subsubsection{S=4}

\begin{small}
\begin{align}
    CS_1 &= \left(
    \begin{tabular}{|c|c|c|}
         \hline
         \quad \quad & \quad \quad & $\hspace{0.07cm} 7\hspace{0.07cm}$  \\
         \hline
         \quad \quad & \quad \quad \\
         \cline{1-2}
         \quad \quad & \quad \quad \\
         \cline{1-2}
    \end{tabular},
    \begin{tabular}{|c|}
         \hline
         $\hspace{0.07cm} \overline{8}\hspace{0.07cm}$  \\
         \hline
    \end{tabular}\right)_C \otimes
    \begin{tabular}{|c|c|c|c|c|c|c|c|}
         \hline
         \quad \quad & \quad \quad & \quad \quad & \quad \quad & \quad \quad & \quad \quad & $\hspace{0.07cm} 7\hspace{0.07cm}$ & $\hspace{0.07cm} \overline{8}\hspace{0.07cm}$   \\
         \hline
    \end{tabular}_S
\end{align}
\end{small}

\subsubsection{S=3}

\begin{small}
\begin{align}
    CS_1 &= \left(
    \begin{tabular}{|c|c|c|}
         \hline
         \quad \quad & \quad \quad & \quad \quad  \\
         \hline
         \quad \quad & \quad \quad \\
         \cline{1-2}
         \quad \quad & $\hspace{0.07cm} 7\hspace{0.07cm}$ \\
         \cline{1-2}
    \end{tabular},
    \begin{tabular}{|c|}
         \hline
         $\hspace{0.07cm} \overline{8}\hspace{0.07cm}$  \\
         \hline
    \end{tabular}\right)_C \otimes
    \begin{tabular}{|c|c|c|c|c|c|c|}
         \hline
         \quad \quad & \quad \quad & \quad \quad & \quad \quad & \quad \quad & $\hspace{0.07cm} 7\hspace{0.07cm}$ & $\hspace{0.07cm} \overline{8}\hspace{0.07cm}$   \\
         \hline
         \quad \quad \\
         \cline{1-1}
    \end{tabular}_S \\
    CS_2 &= \left(
    \begin{tabular}{|c|c|c|}
         \hline
         \quad \quad & \quad \quad & $\hspace{0.07cm} 7\hspace{0.07cm}$  \\
         \hline
         \quad \quad & \quad \quad \\
         \cline{1-2}
         \quad \quad & \quad \quad \\
         \cline{1-2}
    \end{tabular},
    \begin{tabular}{|c|}
         \hline
         $\hspace{0.07cm} \overline{8}\hspace{0.07cm}$  \\
         \hline
    \end{tabular}\right)_C \otimes
    \begin{tabular}{|c|c|c|c|c|c|c|}
         \hline
         \quad \quad & \quad \quad & \quad \quad & \quad \quad & \quad \quad & \quad \quad & $\hspace{0.07cm} 7\hspace{0.07cm}$    \\
         \hline
         $\hspace{0.07cm} \overline{8}\hspace{0.07cm}$ \\
         \cline{1-1}
    \end{tabular}_S \\
    CS_3 &= \left(
    \begin{tabular}{|c|c|c|}
         \hline
         \quad \quad & \quad \quad & $\hspace{0.07cm} 7\hspace{0.07cm}$  \\
         \hline
         \quad \quad & \quad \quad \\
         \cline{1-2}
         \quad \quad & \quad \quad \\
         \cline{1-2}
    \end{tabular},
    \begin{tabular}{|c|}
         \hline
         $\hspace{0.07cm} \overline{8}\hspace{0.07cm}$  \\
         \hline
    \end{tabular}\right)_C \otimes
    \begin{tabular}{|c|c|c|c|c|c|c|}
         \hline
         \quad \quad & \quad \quad & \quad \quad & \quad \quad & \quad \quad & \quad \quad & $\hspace{0.07cm} \overline{8}\hspace{0.07cm}$    \\
         \hline
         $\hspace{0.07cm} 7\hspace{0.07cm}$ \\
         \cline{1-1}
    \end{tabular}_S
\end{align}
\end{small}

\subsubsection{S=2}

\begin{small}
\begin{align}
    CS_1 &= \left(
    \begin{tabular}{|c|c|c|}
         \hline
         \quad \quad & \quad \quad & \quad \quad  \\
         \hline
         \quad \quad & \quad \quad \\
         \cline{1-2}
         \quad \quad & $\hspace{0.07cm} 7\hspace{0.07cm}$ \\
         \cline{1-2}
    \end{tabular},
    \begin{tabular}{|c|}
         \hline
         $\hspace{0.07cm} \overline{8}\hspace{0.07cm}$  \\
         \hline
    \end{tabular}\right)_C \otimes
    \begin{tabular}{|c|c|c|c|c|c|}
         \hline
         \quad \quad & \quad \quad & \quad \quad & \quad \quad & \quad \quad & $\hspace{0.07cm} 7\hspace{0.07cm}$   \\
         \hline
         \quad \quad & $\hspace{0.07cm} \overline{8}\hspace{0.07cm}$ \\
         \cline{1-2}
    \end{tabular}_S \\
    CS_2 &= \left(
    \begin{tabular}{|c|c|c|}
         \hline
         \quad \quad & \quad \quad & \quad \quad  \\
         \hline
         \quad \quad & \quad \quad \\
         \cline{1-2}
         \quad \quad & $\hspace{0.07cm} 7\hspace{0.07cm}$ \\
         \cline{1-2}
    \end{tabular},
    \begin{tabular}{|c|}
         \hline
         $\hspace{0.07cm} \overline{8}\hspace{0.07cm}$  \\
         \hline
    \end{tabular}\right)_C \otimes
    \begin{tabular}{|c|c|c|c|c|c|}
         \hline
         \quad \quad & \quad \quad & \quad \quad & \quad \quad & \quad \quad & $\hspace{0.07cm} \overline{8}\hspace{0.07cm}$   \\
         \hline
         \quad \quad & $\hspace{0.07cm} 7\hspace{0.07cm}$ \\
         \cline{1-2}
    \end{tabular}%
_S
\end{align}
\end{small}

\subsubsection{S=1}

\begin{small}
\begin{align}
    CS_1 &= \left(
    %
_S 
\end{align}
\end{small}

\subsubsection{S=0}

\begin{small}
\begin{align}
    CS_1 &= \left(
    %
_S 
\end{align}
\end{small}

\subsection{$q^5 s^2 \overline{Q}(I=\frac{5}{2})$}

\begin{small}
\begin{align}
    F=\left( 
    \begin{tabular}{|c|c|c|c|c|}
         \hline
         \quad \quad & \quad \quad & \quad \quad & \quad \quad & \quad \quad   \\
         \hline
    \end{tabular}, 
    \begin{tabular}{|c|c|}
         \hline
         $\hspace{0.07cm} 6\hspace{0.07cm}$ & $\hspace{0.07cm} 7\hspace{0.07cm}$  \\
         \hline
    \end{tabular},
    \begin{tabular}{|c|}
         \hline
         $\hspace{0.07cm} \overline{8}\hspace{0.07cm}$  \\
         \hline
    \end{tabular}%
\right)
\end{align}
\end{small}

Since there are two strange quarks, we have to consider the symmetry between them. We use the following notation to represent the symmetry between strange quarks.\\

\begin{small}
$\left(
    %
\right)$
    \end{small}: symmetric between 6 and 7
\\

Similarly, spin can be considered in the same way.

\subsubsection{S=2}

\begin{small}
\begin{align}
    CS_1 &= \left(
    %
_S
\end{align}
\end{small}

\subsubsection{S=1}

\begin{small}
\begin{align}
    CS_1 &= \left(
    %
_S   
\end{align}
\end{small}

\subsubsection{S=0}

\begin{small}
\begin{align}
    CS_1 &= \left(
    %
_S 
\end{align}
\end{small}

\subsection{$q^5 s^2 \overline{Q}(I=\frac{3}{2})$}

\begin{small}
\begin{align}
    F=\left( 
    %
\right)
\end{align}
\end{small}

\subsubsection{S=3}

\begin{small}
\begin{align}
    CS_1 &= \left(
    %
_S
\end{align}
\end{small}

\subsubsection{S=2}

\begin{small}
\begin{align}
    CS_1 &= \left(
    %
_S 
\end{align}
\end{small}

\subsubsection{S=1}

\begin{small}
\begin{align}
    CS_1 &= \left(
    %
_S
\end{align}
\end{small}

\subsubsection{S=0}

\begin{small}
\begin{align}
    CS_1 &= \left(
    %
_S
\end{align}
\end{small}

\subsection{$q^5 s^2 \overline{Q}(I=\frac{1}{2})$}

\begin{small}
\begin{align}
    F=\left( 
    %
\right)
\end{align}
\end{small}

\subsubsection{S=4}

\begin{small}
\begin{align}
    CS_1 &= \left(
    %
_S
\end{align}
\end{small}

\subsubsection{S=3}

\begin{small}
\begin{align}
    CS_1 &= \left(
    %
_S
\end{align}
\end{small}

\subsubsection{S=2}

\begin{small}
\begin{align}
    CS_1 &= \left(
    %
_S 
\end{align}
\end{small}

\subsubsection{S=1}

\begin{small}
\begin{align}
    CS_1 &= \left(
    %
_S 
\end{align}
\end{small}

\subsubsection{S=0}

\begin{small}
\begin{align}
    CS_1 &= \left(
    %
_S
\end{align}
\end{small}

\subsection{$q^4 s^3 \overline{Q}(I=2)$}

\begin{small}
\begin{align}
    F=\left( 
    %
\right)
\end{align}
\end{small}

When there are three or more strange quarks, the number of possible configurations that satisfy the Pauli exclusion principle becomes more complex. When there are two strange quarks, the final symmetry is determined by whether the color and spin are symmetric or antisymmetric. However, when there are three or more strange quarks, even if it initially seems that there is no symmetry in the color or spin, the combined coupling state can be arranged in such a way that it satisfies the symmetric property. Therefore, to represent such cases, the following notation will be used.\\

\begin{small}
$\left(
_S$ \\
\end{small}
    
This state is antisymmetric among 5th, 6th, and 7th quarks for color-spin coupling state, and orthogonal to other $CS_i$ states. Here M means mixed symmetry.

\subsubsection{S=3}

\begin{small}
\begin{align}
    CS_1 &= \left(
    %
_S
\end{align}
\end{small}

\subsubsection{S=2}

\begin{small}
\begin{align}
    CS_1 &= \left(
    %
_S 
\end{align}
\end{small}

\subsubsection{S=1}

\begin{small}
\begin{align}
    CS_1 &= \left(
    %
_S 
\end{align}
\end{small}

\subsubsection{S=0}

\begin{small}
\begin{align}
    CS_1 &= \left(
    %
_S 
\end{align}
\end{small}

\subsection{$q^4 s^3 \overline{Q}(I=1)$}

\begin{small}
\begin{align}
    F=\left( 
    %
\right)
\end{align}
\end{small}

\subsubsection{S=4}

\begin{small}
\begin{align}
    CS_1 &= \left(
    %
_S
\end{align}
\end{small}

\subsubsection{S=3}

\begin{small}
\begin{align}
    CS_1 &= \left(
    %
_S
\end{align}
\end{small}

\subsubsection{S=2}

\begin{small}
\begin{align}
    CS_1 &= \left(
    %
_S 
\end{align}
\end{small}

\subsubsection{S=1}

\begin{small}
\begin{align}
    CS_1 &= \left(
    %
_S  
\end{align}
\end{small}

\subsubsection{S=0}

\begin{small}
\begin{align}
    CS_1 &= \left(
    %
_S 
\end{align}
\end{small}

\subsection{$q^4 s^3 \overline{Q}(I=0)$}

\begin{small}
\begin{align}
    F=\left( 
    %
\right)
\end{align}
\end{small}

\subsubsection{S=3}

\begin{small}
\begin{align}
    CS_1 &= \left(
    %
_S 
\end{align}
\end{small}

\subsubsection{S=2}

\begin{small}
\begin{align}
    CS_1 &= \left(
    %
_S 
\end{align}
\end{small}

\subsubsection{S=1}

\begin{small}
\begin{align}
    CS_1 &= \left(
    %
_S 
\end{align}
\end{small}

\subsubsection{S=0}

\begin{small}
\begin{align}
    CS_1 &= \left(
    %
_S 
\end{align}
\end{small}

\subsection{$s^4 q^3 \overline{Q}(I=\frac{3}{2})$}

For computational convenience, we have arranged the strange quarks at the front. 

\begin{small}
\begin{align}
    F=\left( 
    %
\right)
\end{align}
\end{small}

This case is the same as $q^4 s^3 \overline{Q}(I=2)$. We just need to swap $m_u$ and $m_s$.

\subsection{$s^4 q^3 \overline{Q}(I=\frac{1}{2})$}

In this case, the color-spin coupling states of the 5th, 6th, and 7th quarks do not necessarily need to be antisymmetric. When color and spin are coupled below, it should be considered that the coupling is done in a way that satisfies the given mixed symmetry between the 5th, 6th, and 7th quarks.

\begin{small}
\begin{align}
    F=\left( 
    %
\right)
\end{align}
\end{small}

\subsubsection{S=2}

\begin{small}
\begin{align}
    CS_1 &= \left(
    %
_S
\end{align}
\end{small}

\subsubsection{S=1}

\begin{small}
\begin{align}
    CS_1 &= \left(
    %
_S 
\end{align}
\end{small}

\subsubsection{S=0}

\begin{small}
\begin{align}
    CS_1 &= \left(
    %
_S 
\end{align}
\end{small}

\subsection{$s^5 q^2 \overline{Q}(I=1)$}

\begin{small}
\begin{align}
    F=\left( 
    %
\right)
\end{align}
\end{small}

This case is the same as $q^5 s^2 \overline{Q}(I=\frac{5}{2})$.

\subsection{$s^5 q^2 \overline{Q}(I=0)$}

\begin{small}
\begin{align}
    F=\left( 
    %
\right)
\end{align}
\end{small}

\subsubsection{S=2}

\begin{small}
\begin{align}
    CS_1 &= \left(
    %
_S
\end{align}
\end{small}

\subsubsection{S=1}

\begin{small}
\begin{align}
    CS_1 &= \left(
    %
_S
\end{align}
\end{small}

\subsubsection{S=0}

\begin{small}
\begin{align}
    CS_1 &= \left(
    %
_S 
\end{align}
\end{small}

\subsection{$s^6 q \overline{Q}(I=\frac{1}{2})$}

\begin{small}
\begin{align}
    F=\left( 
    %
\right)
\end{align}
\end{small}

This case is the same as $q^6 s \overline{Q}(I=3)$.

\section*{Acknowledgments}
This work was supported by the Korea National Research Foundation under the grant No. 2023R1A2C300302312


\begin{thebibliography}{99}

\bibitem{Jaffe:1976ig}
R.~L.~Jaffe,
Phys. Rev. D \textbf{15}, 267 (1977)
doi:10.1103/PhysRevD.15.267

\bibitem{Swanson:2006st}
E.~S.~Swanson,
Phys. Rept. \textbf{429}, 243-305 (2006)
doi:10.1016/j.physrep.2006.04.003
[arXiv:hep-ph/0601110 [hep-ph]].

\bibitem{Nielsen:2009uh}
M.~Nielsen, F.~S.~Navarra and S.~H.~Lee,
Phys. Rept. \textbf{497}, 41-83 (2010)
doi:10.1016/j.physrep.2010.07.005
[arXiv:0911.1958 [hep-ph]].

\bibitem{Brambilla:2019esw}
N.~Brambilla, S.~Eidelman, C.~Hanhart, A.~Nefediev, C.~P.~Shen, C.~E.~Thomas, A.~Vairo and C.~Z.~Yuan,
Phys. Rept. \textbf{873}, 1-154 (2020)
doi:10.1016/j.physrep.2020.05.001
[arXiv:1907.07583 [hep-ex]].

\bibitem{Liu:2019zoy}
Y.~R.~Liu, H.~X.~Chen, W.~Chen, X.~Liu and S.~L.~Zhu,
Prog. Part. Nucl. Phys. \textbf{107}, 237-320 (2019)
doi:10.1016/j.ppnp.2019.04.003
[arXiv:1903.11976 [hep-ph]].

\bibitem{Richard:2016eis}
J.~M.~Richard,
Few Body Syst. \textbf{57}, no.12, 1185-1212 (2016)
doi:10.1007/s00601-016-1159-0
[arXiv:1606.08593 [hep-ph]].

\bibitem{Ali:2017jda}
A.~Ali, J.~S.~Lange and S.~Stone,
Prog. Part. Nucl. Phys. \textbf{97}, 123-198 (2017)
doi:10.1016/j.ppnp.2017.08.003
[arXiv:1706.00610 [hep-ph]].

\bibitem{Hosaka:2016pey}
A.~Hosaka, T.~Iijima, K.~Miyabayashi, Y.~Sakai and S.~Yasui,
PTEP \textbf{2016}, no.6, 062C01 (2016)
doi:10.1093/ptep/ptw045
[arXiv:1603.09229 [hep-ph]].

\bibitem{ExHIC:2017smd}
S.~Cho \textit{et al.} [ExHIC],
Prog. Part. Nucl. Phys. \textbf{95}, 279-322 (2017)
doi:10.1016/j.ppnp.2017.02.002
[arXiv:1702.00486 [nucl-th]].

\bibitem{Bicudo:2003rw}
P.~Bicudo and G.~M.~Marques,
Phys. Rev. D \textbf{69}, 011503 (2004)
doi:10.1103/PhysRevD.69.011503
[arXiv:hep-ph/0308073 [hep-ph]].

\bibitem{Park:2017mdp}
A.~Park, W.~Park and S.~H.~Lee,
Phys. Rev. D \textbf{96}, no.3, 034029 (2017)
doi:10.1103/PhysRevD.96.034029
[arXiv:1706.10025 [hep-ph]].

\bibitem{Luo:2022cun}
S.~Q.~Luo, L.~S.~Geng and X.~Liu,
Phys. Rev. D \textbf{106}, no.1, 014017 (2022)
doi:10.1103/PhysRevD.106.014017
[arXiv:2206.04586 [hep-ph]].


\bibitem{Maezawa:2004va}
Y.~Maezawa, T.~Hatsuda and S.~Sasaki,
Prog. Theor. Phys. \textbf{114}, 317-327 (2005)
doi:10.1143/PTP.114.317
[arXiv:hep-ph/0412025 [hep-ph]].

\bibitem{Garcilazo:2016ylj}
H.~Garcilazo and A.~Valcarce,
Phys. Rev. C \textbf{93}, no.3, 034001 (2016)
doi:10.1103/PhysRevC.93.034001
[arXiv:1605.04108 [hep-ph]].

\bibitem{Garcilazo:2016gkj}
H.~Garcilazo,
Phys. Rev. C \textbf{93}, no.2, 024001 (2016)
doi:10.1103/PhysRevC.93.024001

\bibitem{Park:2018ukx}
A.~Park, W.~Park and S.~H.~Lee,
Phys. Rev. D \textbf{98}, no.3, 034001 (2018)
doi:10.1103/PhysRevD.98.034001
[arXiv:1801.10350 [hep-ph]].

\bibitem{Park:2019jff}
A.~Park and S.~H.~Lee,
Phys. Rev. C \textbf{100}, no.5, 055201 (2019)
doi:10.1103/PhysRevC.100.055201
[arXiv:1908.08333 [hep-ph]].

\bibitem{Park:2020qpd}
A.~Park and S.~H.~Lee,
Phys. Rev. D \textbf{102}, no.9, 096024 (2020)
doi:10.1103/PhysRevD.102.096024
[arXiv:2009.06220 [hep-ph]].




\bibitem{Gordillo:2023tnz}
M.~C.~Gordillo and J.~M.~Alcaraz-Pelegrina,
Phys. Rev. D \textbf{108}, no.5, 054027 (2023)
doi:10.1103/PhysRevD.108.054027
[arXiv:2307.08408 [hep-ph]].

\bibitem{Yasui:2009bz}
S.~Yasui and K.~Sudoh,
Phys. Rev. D \textbf{80}, 034008 (2009)
doi:10.1103/PhysRevD.80.034008
[arXiv:0906.1452 [hep-ph]].

\bibitem{DeRujula:1975qlm}
A.~De Rujula, H.~Georgi and S.~L.~Glashow,
Phys. Rev. D \textbf{12}, 147-162 (1975)
doi:10.1103/PhysRevD.12.147

\bibitem{Oka:1981ri}
M.~Oka and K.~Yazaki,
Prog. Theor. Phys. \textbf{66}, 556-571 (1981)
doi:10.1143/PTP.66.556

\bibitem{Park:2019bsz}
A.~Park, S.~H.~Lee, T.~Inoue and T.~Hatsuda,
Eur. Phys. J. A \textbf{56}, no.3, 93 (2020)
doi:10.1140/epja/s10050-020-00078-z
[arXiv:1907.06351 [hep-ph]].

\bibitem{Noh:2024nvp}
S.~Noh, A.~Park, H.~Yun, S.~Cho and S.~H.~Lee,
Phys. Lett. B \textbf{862}, 139278 (2025)
doi:10.1016/j.physletb.2025.139278
[arXiv:2408.00715 [hep-ph]].

\bibitem{Glozman:1995fu}
L.~Y.~Glozman and D.~O.~Riska,
Phys. Rept. \textbf{268}, 263-303 (1996)
doi:10.1016/0370-1573(95)00062-3
[arXiv:hep-ph/9505422 [hep-ph]].






\end{thebibliography}
\end{document}